\documentclass[a4paper,11pt]{article}
\usepackage{jheppub} 

\usepackage{bm,amsmath,amssymb,slashed,graphicx,%
            enumerate,alltt,xspace,multirow,xcolor,mathrsfs}
\usepackage{fancyvrb}
\usepackage{cancel}

\usepackage{subcaption}

\usepackage{changes}

\definechangesauthor[name=pn, color=red]{pn}

\RequirePackage{braket}
\RequirePackage{graphicx}

\RequirePackage[numbers,sort&compress]{natbib}

\makeatletter
\g@addto@macro\bfseries{\boldmath}
\makeatother

\definecolor{labelkey}{rgb}{0,0.5,0.0}

\usepackage{listings}
\definecolor{darkgreen}{rgb}{0,0.4,0}

\newcommand\bk{b_k}

\newcommand\mathd{\mathrm{d}}

\newcommand{\as}{\alpha_s}

\newcommand{\POWHEG}{{\tt POWHEG}}

\newcommand{\POWHEGhvq}{{\tt POWHEG-hvq}}
\newcommand{\POWHEGnloDec}{{\tt POWHEG-ttb\_NLO\_dec}}
\newcommand{\POWHEGbbfourl}{{\tt POWHEG-b\_bbar\_4l}}
\newcommand{\ttMiNNLO}{{\tt MiNNLO-ttbar}}
\newcommand{\Pythiaeight}{{\tt Pythia8{}}}

\newcommand{\Cf}{C_{\rm \scriptscriptstyle F}}

\newcommand\Chel{C_{\rm hel}}
\newcommand\mtt{M_{t{\bar t}}}

\title{Spin Correlations in $t{\bar t}$ Production and Decay\\ at the LHC
  in QCD Perturbation Theory.}

\preprint{
  \begin{flushright}
  LAPTH-017/25  
  \end{flushright}
}

\author[b]{Paolo Nason,}
\author[a,b,1]{Emanuele Re\note{On leave of absence from LAPTh, Universit\'e Grenoble Alpes, Universit\'e Savoie Mont Blanc, CNRS,
F-74940 Annecy, France.},}
\author[a,b]{Luca Rottoli}

\emailAdd{paolo.nason@mib.infn.it}
\emailAdd{emanuele.re@mib.infn.it}
\emailAdd{luca.rottoli@unimib.it}

\affiliation[a]{Dipartimento di Fisica G. Occhialini, Universit\`a degli Studi di Milano-Bicocca}
\affiliation[b]{INFN, Sezione di
  Milano-Bicocca, Piazza della Scienza 3,20126 Milano, Italy}

\date{Received: date / Accepted: \today}

\abstract{ In this work we consider the QCD predictions for spin
  correlations in $t\bar{t}$ production in hadronic collisions. In
  view of recent tensions between experimental data and theoretical
  calculations, it has been argued that one should include in the
  predictions also the effects of the production of the $\eta_t$,
  i.e. the pseudoscalar $t\bar{t}$ bound state, or alternatively
  the full effects of the non-relativistic dynamics of the
  $t\bar{t}$ pair near threshold.  This implies the resummation of all
  corrections that scale like powers of $\as/v$ (where $v$ is the
  velocity of the top quark in the $t\bar{t}$ rest frame) which are
  dominated by values of $v$ of order $\as$. In this work, we show that,
  since the
  observables that are usually considered for these studies are
  integrated cross sections up to a $t\bar{t}$ mass cut that is not small,
  it is possible to perform the calculation using perturbation theory,
  considering only the contributions that scale as the first few
  powers of $\as/v$.  We examine the implications of our approach by computing
  corrections to nominal Monte Carlo results for correlation-sensitive
  observables, and compare them with available data, showing that the
  tension with data is no longer present.}


\begin{document}

\maketitle


\section{Introduction}
\label{sec:intro}
The top spin in top leptonic decays is strongly correlated with the direction
of the outgoing antilepton. More specifically, in the rest frame of the decaying top, the top spinor
is the eigenstate with positive eigenvalue of
$\vec\sigma \cdot \vec{l}$, where $l$ is the direction of the antilepton
in the top rest frame, and $\vec{\sigma}$ are the Pauli sigma matrices.
Moreover, the production of a $t\bar{t}$ pair near threshold has a sizeable spin singlet
component, so that quantum correlations of the spins of the top and antitop are large enough to establish a violation of the Bell inequalities~\cite{Afik:2020onf,Fabbrichesi:2021npl,Severi:2021cnj,Afik:2022kwm,Aoude:2022imd,Aguilar-Saavedra:2022uye,Fabbrichesi:2022ovb,Severi:2022qjy,Dong:2023xiw,Aguilar-Saavedra:2023hss,Duch:2024pwm,Maltoni:2024tul,Aguilar-Saavedra:2024hwd,Barr:2024djo,Cheng:2024btk}. Both the ATLAS~\cite{ATLAS:2023fsd} and the CMS~\cite{CMS:2024zkc,CMS:2024pts,CMS:2025kzt} collaborations
have observed this phenomenon. It turns out, however, that the amount of correlation that they measure
is larger than the one computed with standard Monte Carlo generators.
In order to alleviate this discrepancy it has been argued that one should add the contribution
of the $t\bar{t}$ pseudoscalar bound state $\eta_t$~\cite{Fuks:2021xje,Severi:2021cnj,Maltoni:2024tul,Aguilar-Saavedra:2024mnm}. This state is characterised by top velocities of order $\as$,
and a binding energy of order $\as^2 m$, where $m$ is the top mass. The binding energy turns out
to be of the same order of the top width $\Gamma_t$, so that, by the uncertainty principle, we can conclude that no
bound state is formed. Nevertheless, a small
bump in the production cross section should be visible if one had the experimental resolution to measure precisely the
mass of the $t{\bar t}$ pair, or if one was looking at production channels where the mass is well-constrained \cite{Hoang:1998xf,Hoang:2013uda,Beneke:2015kwa}, or at
final states arising from the annihilation of the $t\bar{t}$ pairs into particles that can be well measured~\cite{Dugad:2016kdv,Kawabata:2016aya}.
Attempts to include the effects of the $\eta_t$ have been carried out
by the experiments themselves~\cite{CMS:2024pts,CMS:2025kzt}, and also in
refs.~\cite{Fuks:2021xje,Maltoni:2024tul,Fuks:2024yjj,Aguilar-Saavedra:2024mnm}. To
improve the description of the region close to threshold, the
all-order resummation of non-relativistic effects in $t\bar{t}$
production has been subject of various
studies, see for
example~\cite{Fadin:1987wz,Fadin:1988fn,Fadin:1990wx,Hagiwara:2008df,Kiyo:2008bv,Sumino:2010bv,%
  Beneke:2011mq,Beneke:2012wb,Ju:2020otc,Garzelli:2024uhe}. These
effects are typically not included in state-of-the-art predictions for
differential top-quark pair production, which nowadays reach NNLO
accuracy in
QCD~\cite{Czakon:2015owf,Czakon:2016ckf,Catani:2019hip,Catani:2020tko}
and can be supplemented by the inclusion of electroweak (EW) corrections
\cite{Czakon:2017wor} and by the resummation of soft and small-mass
logarithms~\cite{Czakon:2018nun,Czakon:2019txp}.

\begin{figure}[htb]
  \begin{center}
    \includegraphics[width=0.4\textwidth]{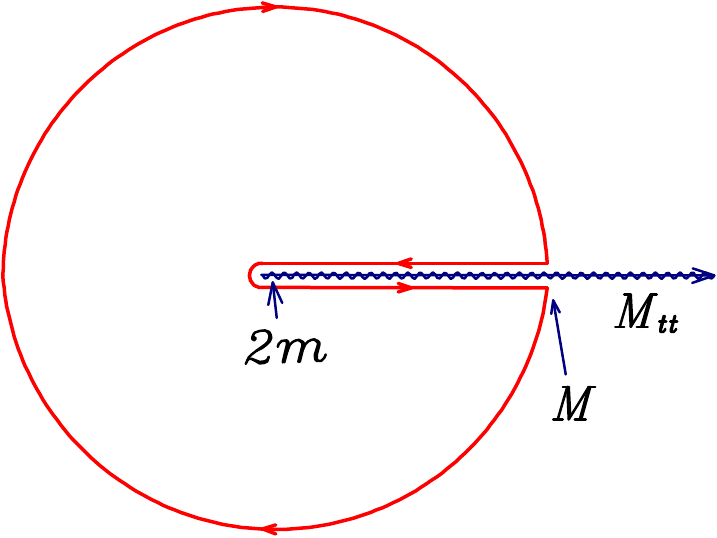}
  \end{center}
  \caption{\label{fig:contour} Illustration of the cross section integrated
    in the invariant mass of the $t\bar{t}$ pair up to
    a given cut~$M_{t\bar{t}}^{(\rm cut)}$ in terms of a
    contour integral of the forward amplitude for complex energies.}
\end{figure}

The motivation of the present work arises from the observation that, since we cannot
fully resolve the invariant mass of the $t\bar{t}$ system
experimentally, it should not be necessary to include the full resummation of
non-relativistic effects.  The
experimental collaborations typically deal with integrated
distribution in the $t\bar{t}$ mass up to a mass cut
$M$ of the order of $380-400$~GeV.
Intuitive reasoning leads us to conclude that under these
circumstances enhanced contributions of order $(\as/v)^n$ (where $v$ is the top velocity in the $t \bar t$ rest frame) which arise to all order in
perturbation theory should be sensitive to the velocities $v$
corresponding to the adopted mass cut $M$ rather than $v \sim \as$,  which would require full resummation and may also give rise to bound state formation.
 This follows from the fact that we are dealing with a cross section for a final state in
an $s$-wave angular momentum state, which is
proportional to the squared amplitude integrated over the final state
kinematics.
By the optical theorem, the cross section can be written as the
imaginary part of the forward amplitude, and thus the integral of the cross section
over the energy of the $t\bar{t}$ system up to a given cut
$M$ can be expressed as a contour integral
in the (complex) energy plane around the real
axis, where the bound state poles and the cut corresponding to the
production of free tops are located. This contour integral can be deformed
into an integral around a circle with a radius equal to $M-2 m$, where $m$ is the top mass,
as illustrated in Fig.~\ref{fig:contour},
which shows that the integration region can only be sensitive to
energies that are far from threshold by an amount of order $M-2m$.
One thus expects that the integrated cross
section will acquire the structure of $1/v$ singularities,
where $v$ is the velocity of the top in the $t\bar t$ rest
frame when the mass of the pair is equal to $M$. Since
$v \gg \as$, the integrated cross section can be expanded
into a convergent series of the coupling constant, and no resummation
is needed.

The finite width of the top introduces yet another smearing of the
threshold cross section. This smearing is essential when describing
the shape of the cross section as a function of the invariant mass
near threshold~\cite{Fadin:1987wz,Fadin:1988fn,Fadin:1990wx},
in the cases when this can be measured with
high precision. If instead the resolution in $m_{t{\bar t}}$
is much larger than the top width, this smearing is irrelevant.

The observation that one should not add the bound state contributions
to cross sections involving integrated spectral densities is not new,
having appeared in several instances in the past. It appeared in
ref.~\cite{Beneke:2016jpx} in the context of threshold enhanced
contributions to the total cross section in $t \bar t$ hadronic production. In
ref.~\cite{Melnikov:2014lwa} it was demonstrated that, contrary to
some suggestions in the literature, one should not add the
contribution of positronium states to the photon vacuum polarisation
in the computation of the electron $g-2$ (see also ref.~\cite{Eides:2014swa}).  In the same reference it is
argued that (again contrary to the suggestion of some authors) the
contribution of the $t{\bar t}$ bound states should not be added to
vacuum polarisation calculations affecting electroweak
observables. Further references on the argument can be found
in~\cite{Braun:1968njz, Novikov:1977dq, Voloshin:1979uv,
  Voloshin:1984zzn, Smith:1994ev}, where the 1968 work of M.~Braun
seems to be the oldest one where this observation appears. These
references were pointed out to us by M. Beneke and K. Melnikov after
we completed our analysis of the problem.\footnote{ We thank M. Beneke
  and K. Melnikov for pointing out these results to us.} In this work
we present our approach, that seems to us particularly simple.  In
view of the fact that apparently the results of
refs.~\cite{Beneke:2016jpx, Melnikov:2014lwa, Eides:2014swa, Braun:1968njz,
  Novikov:1977dq, Voloshin:1979uv, Voloshin:1984zzn, Smith:1994ev}
have not yet become common knowledge in the theoretical physics
community, we hope that our contribution will help in this respect.

In our approach we first consider the simplest quantum
mechanical system that exhibits a bound state, i.e. the problem of a
single non-relativistic particle in one space dimension, in the
presence of a (negative) delta-function potential. This model
illustrates what happens when we consider the integral of the spectrum
up to a given energy cut far
above the bound state energy, and the structure of the perturbative
expansion in term of inverse powers of the velocity associated with
the energy cut. This simple model captures all the
features of the problem, so that when we consider our case of interest
no further complications do arise.

The paper is organised as follows. In Section~\ref{sec:spincorr} we summarise the main features
of the spin correlations in $t\bar{t}$ production near threshold, and why and how to improve
this description by adding non-relativistic Coulomb interactions.
In Section~\ref{sec:toymodel} we describe a toy model that clarifies the interplay between bound
state effects and production of freely propagating states in the perturbative description
of the production process. In Section~\ref{sec:realcase} we generalise the result obtained
with the toy model to the full realistic case. Section~\ref{sec:modeling} deals
with the modelling of spin correlations in hadronic collisions, and introduce the main observables
used to study correlation. In Section~\ref{sec:anatomy} we show the result of the calculations of correlation
observables at the Born level. By comparing it with results obtained with full simulations
it is shown that the Born level result gives already a good description of the correlations.
In the same Section we show results for the Born level predictions augmented with the addition of the
leading threshold corrections up to relative order $\as^3$, as we computed in
Section~\ref{sec:realcase}. In Section~\ref{sec:nominal} we show results obtained with
the nominal Monte Carlo generators
\POWHEGhvq{} \cite{Frixione:2007nw}, \POWHEGnloDec{} \cite{Campbell:2014kua},
and \POWHEGbbfourl{} \cite{Jezo:2016ujg}, in
order to assess the effect of using implementations of correlations in decay with
increasing accuracy. In Section~\ref{sec:nloimproved} we show how to improve the
results of the nominal NLO generators by adding the threshold enhanced corrections
computed in this work, and show the comparison with available data. In Section~\ref{sec:minnlo}
we consider the NNLO generator \ttMiNNLO{} \cite{Mazzitelli:2020jio,Mazzitelli:2021mmm},
explain how to improve it with threshold
effects, and present predictions and comparison with data. In Section~\ref{sec:conclusions}
we present our conclusions.

\section{Spin correlations in $t\bar{t}$ production near threshold}%
\begin{figure}[htb]
  \begin{center}
    \includegraphics[width=0.25\textwidth]{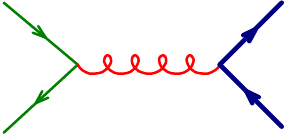} \hskip 2cm
    \includegraphics[width=0.4\textwidth]{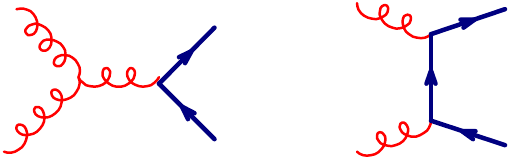}
  \end{center}
  \caption{\label{fig:ttprod} Diagrams involved in heavy quark production at leading order.}
\end{figure}

\label{sec:spincorr}
In this section we summarise the main features, and collect the main
formulae, of $t\bar{t}$ production and decay near threshold.  At
leading order, the production process proceeds via quark-antiquark
annihilation and gluon-gluon fusion, according to the diagrams
reported in fig.~\ref{fig:ttprod}.
Production near threshold implies that the zero orbital angular momentum prevails. Thus,
the total final state angular momentum is determined solely by the spins of the top quarks. In $q\bar{q}$
annihilation the $t\bar{t}$ pair must be in a spin triplet configuration, since the production goes through
a single virtual gluon. In the $gg$ channel, on the other hand, the $t\bar{t}$ pair must be in a spin singlet
configuration. This is a consequence of the Landau-Yang theorem~\cite{Landau:1948kw,Yang:1950rg},
which forbids the coupling of a spin one system to a pair of massless
spin one particles, and of the fact that the  $t\bar{t}$ system can only be in a singlet or triplet spin state.
The $q{\bar q}$ production
channel is characterised by a $t\bar{t}$ pair in a colour octet state, while $gg$ production includes
both a colour singlet and a colour octet component.
The spin and colour averaged squared amplitudes, separated by colour channels, are given by
\begin{eqnarray}
  h^{(0)}_{q\bar{q}}&=& \frac{C_F^2}{D_A}(2\tau_1^2+2\tau_2^2+\rho), \label{eq:hqq} \\
  h^{(0)}_{gg} &=& \frac{2T_f}{D_A}\left(\left[\frac{1}{D_A}\frac{C_F}{\tau_1\tau_2}\right]_{\rm sng}
                   +\left[\frac{D_A-1}{D_A}\frac{C_F}{\tau_1\tau_2}-C_A\right]_{\rm oct} \right)
                   \left(\tau_1^2+\tau_2^2+\rho-\frac{\rho^2}{4\tau_1\tau_2}\right), \label{eq:hgg} 
\end{eqnarray}
where the separation of the singlet and octet component in $h^{(0)}_{gg}$ has been clearly shown. We have followed the notation
of ref.~\cite{Nason:1989zy}. The $h$ functions represent the spin and colour summed and averaged
squared amplitudes for the corresponding processes, where the $\as^2/m^2$ factor has been removed.
Defining $p_{1/2}$ the incoming momenta and $k_{1/2}$ the outgoing ones, we have
$\rho=4m^2/(2p_1\cdot p_2)$, $\tau_{1/2}=p_1\cdot k_{1/2}/p_1\cdot p_2$, with  $\tau_1+\tau_2=1$, $D_A=8$ is the dimensionality of the adjoint representation, $T_f=1/2$ and $C_F=4/3$.
The corresponding total cross sections are given by
\begin{eqnarray}
  \sigma_{q{\bar q}}&=&\sigma_{q{\bar q}}^{\rm (8)}=\frac{\as^2}{m^2}\frac{\pi v \rho}{27}[2+\rho]\approx \frac{\as^2}{m^2}\frac{\pi v}{9}, \\
  \sigma_{gg}&=&\sigma_{gg}^{\rm (1)}+\sigma_{gg}^{\rm (8)}, \\
  \sigma_{gg}^{\rm (1)}&=&\frac{\as^2}{m^2} \frac{\pi v \rho}{384}\left[\frac{1}{v}\log\frac{1+v}{1-v}(4+4\rho-2\rho^2)-4-4\rho\right]
                           \approx \frac{\as^2}{m^2} \frac{\pi v}{192}[2], \\
  \sigma_{gg}^{\rm (8)}&=&\frac{\as^2}{m^2} \frac{\pi v \rho}{384}\left[\frac{1}{v}\log\frac{1+v}{1-v}(28+28\rho+4\rho^2)-52-58\rho\right]
                                                    \approx \frac{\as^2}{m^2} \frac{\pi v}{192}[5],
\end{eqnarray}                        
where the $\approx$ equality refers to the threshold limit.

We will be interested in Coulombic corrections to the final state $t{\bar t}$ pair. Decomposing the $g g$ Born amplitude into
a singlet and octet component, we can write
\begin{equation}
  {\cal M}^{(gg)}_{i j} =  {\cal M}^{(gg,8)}_{i j}+{\cal M}^{(gg,1)}_{i j} = \left({\cal M}^{(gg)}_{i j} - \frac{1}{N_c}M_{ii} \delta_{ij} \right) + \frac{1}{N_c}M_{ii} \delta_{ij}.
\end{equation}
The colour structure of a Coulombic interaction in the final state can be immediately obtained using the formula
\begin{equation}
  T^a_{i_1 i_1'} T^a_{i_2' i_2}=\frac{1}{2}\left(\delta_{i_1i_2}\delta_{i_1'i_2'}-\frac{1}{N_c}\delta_{i_1i_1'}\delta_{i_2i_2'}\right),
\end{equation}
yielding
\begin{eqnarray}
  T^a_{i_1 i_1'}  {\cal M}^{(gg,8)}_{i_1'i_2'} T^a_{i_2' i_2}
  &=&-\frac{1}{2N_C}  {\cal M}^{(gg,8)}_{i_1 i_2}, \label{eq:octectcoup}\\
  T^a_{i_1 i_1'}  {\cal M}^{(gg,1)}_{i_1'i_2'} T^a_{i_2' i_2}
  &=&\frac{1}{2N_C}   {\cal M}^{(gg)}_{ii}\left(\delta_{i_1i_2}N_c-\frac{1}{N_c}\delta_{i_1i_2}\right)=C_F  {\cal M}^{(gg,1)}.
       \label{eq:singletcoup}
\end{eqnarray}
We thus have an attractive $C_F\as$ coupling for the singlet, and a
repulsive $-\as/(2N_c)$ one for the octet.

In summary, there are mechanisms (i.e. the Landau-Yang theorem) that
enhance spin correlations in $t{\bar t}$ production near threshold at
the LHC, where gluon fusion prevails. Furthermore, Coulombic
interactions can enhance/deplete the cross section near threshold,
depending upon the colour channels.  The experimental collaborations
impose limits on the mass of the $t{\bar t}$ system to enhance the
correlation signal.  Thus it becomes important to asses reliably the
threshold enhanced corrections.
\subsection{Radiative corrections near threshold}\label{sec:powercount}
It is well known that there are enhanced perturbative corrections in
top pair production near threshold. One finds corrections of relative
order $\as/v$ at NLO, and $(\as/v)^2$ at NNLO. Terms of order
$(\as/v)^3$ are actually absent, but there are corrections localised
very near threshold that contribute at order $\as^3$.  The Born cross
section near threshold is of order $\as^2 v/m^2$, so that the enhanced
NLO and NNLO corrections behave as $\as^3/m^2$ and $\as^4/(v m^2)$.
Intuitive reasoning leads to conclude that bound state corrections
contribute at order $\as^5$, i.e. N$^3$LO. This follows from the fact
that bound state production, besides including an $\as^2$ factor for
the hard production process, should also include an extra $\as^3$
factor arising from the square of the wave function at the origin of
the Coulombic system formed by the $t{\bar t}$ pair.\footnote{ In a
  Coulombic system the Bohr radius is of order $1/(m\as)$, and the
  square of the wave function at the origin is of the order of the inverse of the
  cube of the Bohr radius.}  From dimensional analysis we thus
conclude that the bound state contribution has the form
$\as^5 \delta(E-2m)/m$ where $2m$ is equal to the bound state energy
up to corrections of order $\as^2$. The integral in the energy up to a
given energy cut $E_c$ (relatively near threshold) for the Born, NLO,
NNLO and N$^3$LO contributions are of order
\begin{equation}
  {\rm Born} \approx \frac{\as^2 v^3}{m},\;
  {\rm\scriptstyle NLO} \approx \frac{\as^2 v^3}{m}\left(\frac{\as}{v}\right),\;
  {\rm\scriptstyle NNLO} \approx \frac{\as^2 v^3}{m}\left(\frac{\as}{v}\right)^2,\;
  {\rm\scriptstyle N^3LO}\approx \frac{\as^2 v^3}{m}\left(\frac{\as}{v}\right)^3\,,
\end{equation}
where now $v\approx \sqrt{(E_c-2m)/m}$
(notice that the N$^3$LO correction does not depend upon the cut, as
appropriate for the integral of a $\delta$ function).
This suggests that, to all
orders in perturbation theory, corrections of relative order
$(\as/v)^n$ to the integrated cross section
should arise, yielding a convergent perturbative
expansion provided $v$ is not too small. When $v \approx \as$, on
the other hand, the series diverges, and it should be fully resummed
in order to obtain sensible results.
\subsection{The non-relativistic limit}
Top production dynamics near threshold can be studied in terms of non-relativistic
quantum mechanics. In the non-relativistic limit, the production scale of the
order of $1/m$ is very small, and can be assimilated to a point. We thus consider
the creation of a $t$ and ${\bar t}$ quarks at the same point. The corresponding
two body problem can be reduced as usual to the problem of a single particle
with a reduced mass $m/2$, created at the origin of a static Coulomb field.
The solution of the quantum-mechanical problem is embodied by the Green's function
$G(t,\vec{x},0,\vec{x}_0)$, which is the amplitude for a particle localised at the
point $\vec{x}_0$ at time zero to be found at point $\vec{x}$ at time $t$.
It satisfies the Schr\"odinger equation
\begin{equation}
  \left[i\frac{\partial}{\partial t}-H\right]G(t,\vec{x},0,\vec{x}_0)
  =i\delta(t)\delta^3(\vec{x}-\vec{x_0}),
\end{equation}
where in our units ${\hbar}$ equals one.
Taking the Fourier transform in time and assuming that $G$ vanishes for negative $t$
we obtain
\begin{equation}
  (E+i\epsilon -H) R(E,\vec{x},\vec{x}_0)=\delta^3(\vec{x}-\vec{x_0}),
\end{equation}
where $R$ (or resolvent) is the time Fourier transform of $G$. The sign
of $\epsilon$ is the appropriate one for a retarded Green's function.
Since we are interested in configurations that are dominant in the threshold
limit, we only consider $s$-wave final states. Thus, the integration over the
final state kinematics is a simple factor, and the cross section for our process, using the
optical theorem, is given by the imaginary part of the forward amplitude
$R(E+i\epsilon,\vec{x}_0,\vec{x}_0)$, where $\vec{x}_0$ is the origin of the Coulomb
potential. The resolvent, in operator notation is given by
\begin{equation}
  R(E)=\frac{1}{E-H},
\end{equation}
and is an analytic function of $E$ up to singularities for real negative $E$ due to
bound states, and to a cut for $E$ real and positive, due to freely propagating states.
Since the cross section is proportional to the imaginary part of the forward
resolvent,
we expect that its integral
up to a sufficiently large energy cut can also be expressed
as an integral of the forward resolvent along a circle in the complex
energy plane that is far from the
origin (see Fig.~\ref{fig:contour}). Thus
we expect that the $1/v$ enhanced corrections
will be characterised by a relatively large $v$, therefore giving rise
to a convergent perturbative expansion.
In order to explore this feature in some detail,
in the following section we discuss a
simple quantum mechanical model, i.e. the case of a single particle in a delta
function potential in one dimension. This example is sufficient to
clarify the problem, and the generalisation to the full $t\bar{t}$
production process will be straightforwardly obtained following the same line
of reasoning.

\section{The toy model}\label{sec:toymodel}
We consider a single particle of mass $m$ in a potential
$V(x)=-\lambda\delta(x)$. The Schr\"odinger equation is
\begin{equation}
  \left(-\frac{1}{2m}\frac{\mathd^2}{\mathd x^2}-\lambda \delta(x)\right)\psi = E\psi,
\end{equation}
that is easily solved for the (normalised) eigenstates
\begin{eqnarray}
  \psi_b(x) &=& \theta(\lambda)
     \sqrt{k_b}\left[e^{k_b x}\theta(-x)+e^{-k_b x}\theta(x)\right], \label{eq:psibound}\\
  \psi_k(x) &=& \sqrt{\frac{2}{L(1+\bk^2)}}\left[\cos(kx)-\frac{x}{|x|}\bk\sin(kx)\right], \label{eq:psicont}\\
  \hat\psi_k(x) &=& \sqrt{\frac{2}{L}} \sin(kx),\label{eq:psiodd}
\end{eqnarray}
where the suffix $b$ denotes the bound state, while $k$ denotes the
absolute value of the momentum of freely propagating states. We have
$k_b=m\lambda$, $\bk=m\lambda/k$, and the bound state energy is
$E_b=-m\lambda^2/2$. We denote with $L$ the (large) size of the
system. In the following we ignore the parity-odd solutions
$\hat{\psi}$, since they propagate freely and play no role here.

We remark that if $\lambda<0$ the solution in eq.~(\ref{eq:psibound})
will be absent, yielding a function that is not normalizable.
The presence of the factor $\theta(\lambda)$ in eq.~(\ref{eq:psibound}) allows us to
treat simultaneously also the case of $\lambda<0$ in the following manipulations.

Since we know the spectrum we can easily compute the resolvent as
\begin{equation}
  R(E,x_1,x_2)=\frac{\psi_b(x_1)\psi_b(x_2)}{E-E_b}+\sum_k \frac{\psi_k(x_1)\psi_k(x_2)}{E-E_k},
\end{equation}
where $E_k=k^2/(2m)$.
Replacing as usual $\sum_k\to L/(2\pi)\int \mathd k$, we obtain the
forward Green's function
\begin{equation}
  R(E,0,0)=\theta(\lambda)
  \frac{k_b}{E-E_b}+\frac{1}{\pi}\int_0^\infty \mathd k \frac{1}{1+\bk^2}\frac{1}{E-E_k},
\end{equation}
whose imaginary part is proportional to the spectral density
\begin{equation}\label{eq:rhomodel}
  \rho(E)=\theta(\lambda)k_b \delta(E-E_b)+\frac{m}{\pi k}\frac{1}{1+\bk^2},
\end{equation}
where we imply that $k=\sqrt{2mE}$, and in the second term we assume an implicit $\theta(E)$ factor.

If we expand $\rho(E)$ in powers of the coupling $\lambda$
we obtain powers of $\lambda/v$, where $v=k/m$ is the velocity of the particle,
in full analogy with the case of the $\alpha/v$ singularities
arising in the Coulomb problem.
Because of these singularities, however, the coefficients of the
expansion are not integrable in the small energy region
starting at order $\lambda^2$
\begin{equation}
  \rho(E) = \frac{m}{\pi k} + \theta(\lambda)\lambda m  \delta(E)
  - \lambda^2 \frac{m}{\pi k} \left(\frac{m}{k}\right)^2+\ldots,
\end{equation}
(where we have replaced $\delta(E-E_B) \to \delta(E)$ neglecting terms
arising from the Taylor expansion of the $\delta$ function, which start contributing
at order $\lambda^3$).  However, if we first perform
the $E$ integration up to some (positive) upper limit $E_{\rm cut}$
we get a finite result,
and expanding it in $\lambda$ we get
\begin{eqnarray}
  \int_{-\infty}^{E_{\rm cut}} \mathd E \rho(E)
  &=& \frac{1}{\pi} k_{\rm cut}
      +\theta(\lambda) \lambda m  \delta(E)
      - \lambda \frac{1}{\pi} m \arctan\frac{k_{\rm cut}}{m\lambda}\nonumber \\
  &=&\frac{1}{\pi} k_{\rm cut}
  + \lambda m \left[\theta(\lambda)-\frac{|\lambda|}{\lambda}\frac{1}{2}\right]+
  \lambda^2 \frac{m}{\pi} \frac{m}{k_{\rm cut}} + \ldots \label{eq:intrhoexp},
\end{eqnarray}
(where $k_{\rm cut}=\sqrt{2mE_{\rm cut}}$)
that is a sensible expansion as long as $m\lambda/k_{\rm cut}$ is small.
Notice that
\begin{equation}
  \theta(\lambda)-\frac{|\lambda|}{\lambda}\frac{1}{2}=\frac{1}{2}
\end{equation}
so that we have
\begin{equation}
   \int_{-\infty}^{E_{\rm cut}} \mathd E \rho(E)=\frac{1}{\pi} k_{\rm cut}
  + \frac{\lambda}{2} m +
  \lambda^2 \frac{m}{\pi} \frac{m}{k_{\rm cut}} + \ldots,
\end{equation}
irrespective of the sign of $\lambda$.
This must be the
case, since the sign of the coupling should not affect the coefficients
of a valid perturbative expansion.

The result just obtained supports the intuitive argument given
earlier. The integral of the spectral density can be related to a
contour integral of the resolvent in the complex plane, with
$|E|=E_{\rm cut}$, in a region far away from threshold, where an
expansion of the resolvent in powers of $\lambda$ is certainly
valid. We have thus learned that although the perturbative expansion
of $\rho(E)$ seems to be ill-defined, its integral up to a given
energy cut much larger than $m\lambda^2$ has a well defined
perturbative expansion, suggesting that a perturbative expansion for
$\rho(E)$ should also be possible as long as we interpret its
coefficients as distributions.

According to the analyticity argument given earlier, also the integral
of $\rho(E)$ multiplied by an arbitrary analytic function with radius
of convergence larger than a given cut $E_{\rm cut}$ should have a
definite perturbative expansion. In order to prove it, it is enough to
demonstrate that for $n\ge 0$
the integral
\begin{equation}
  I_n=  \int_{-\infty}^{E_{\rm cut}}  \mathd E E^{n} \rho(E)
\end{equation}
has a well-defined perturbative expansion for any $n$, as long as
$E_{\rm cut}>\lambda^2 m/2$. In fact, we find
\begin{eqnarray}
  I_n
  &=& \theta(\lambda) k_bE_b^n
      +\frac{1}{\pi}\int_0^{k_{\rm cut}}\mathd k \frac{E^n}{1+\bk^2}\nonumber \\
  &=& \theta(\lambda) k_bE_b^n
      +\frac{1}{\pi}\int_0^{k_{\rm cut}}\mathd k E^n\left[\frac{1}{1+\bk^2}
      -\sum_{i=0}^n(-\bk^2)^i\right]
      +\frac{1}{\pi}\int_0^{k_{\rm cut}}\mathd k E^n \sum_{i=0}^n(-\bk^2)^i\phantom{aaaaaaaa}
      \label{eq:InToy}
\end{eqnarray}
where as before $b_k=m\lambda/k$.
The subtractions in the square bracket are such that the middle integral
remains finite also in the limit $k_{\rm cut}\to\infty$, and at the same time
no divergences for $E\to 0$ are introduced, so that also the last integral
is well-defined.  The second term can be further manipulated to give
\begin{eqnarray}
  && \frac{1}{\pi}\int_0^{k_{\rm cut}}\mathd k E^n\left[\frac{1}{1+\bk^2}
     -\sum_{i=0}^n(-\bk^2)^i\right]=-\frac{1}{\pi}\int_0^{k_{\rm cut}}\mathd k E^n
     \left[(-)^n\frac{\bk^{2n}}{1+\bk^{-2}} \right] \nonumber \\
  &=& -\frac{E_b^n}{\pi}\int_0^{k_{\rm cut}}\mathd k
     \frac{1}{1+\bk^{-2}}  =  -\frac{E_b^n}{\pi}\int_0^\infty\mathd k
     \frac{1}{1+\bk^{-2}} + \frac{E_b^n}{\pi}\int_{k_{\rm cut}}^\infty \mathd k
      \frac{1}{1+\bk^{-2}} \nonumber \\
  &=& - \frac{1}{2}|\lambda|m E_b^n  + \frac{E_b^n}{\pi}\int_{k_{\rm cut}}^\infty \mathd k
      \frac{1}{1+\bk^{-2}},\label{eq:InToy2}
\end{eqnarray}
where we have used the identity $(-E)^n \bk^{2n}=E_b^n$. As before, the first term
of eq.~(\ref{eq:InToy}) combines with the first term of eq.~(\ref{eq:InToy2}) as
\begin{equation}
  \theta(\lambda) k_b E_b^2 - \frac{1}{2}|\lambda|m E_b^n=\frac{1}{2} k_b E_b^2,
\end{equation}
so that at the end
\begin{eqnarray}
  I_n&=& \frac{1}{2} k_bE_b^n
         -\frac{E_b^n}{\pi}\int_{k_{\rm cut}}^{\infty}\mathd k \sum_{i=1}^\infty (-\bk^2)^i
        + \frac{1}{\pi}\int_0^{k_{\rm cut}}\mathd k E^n \sum_{i=0}^n(-\bk^2)^i, \label{eq:In}
\end{eqnarray}
where all integrals are convergent order by order in perturbation
theory in $\lambda$, and the infinite sums are all convergent, as long
as our assumption $E_{\rm cut}>m\lambda^2/2$ holds.

Notice that our result can be summarised as
the replacement
\begin{equation}
  \rho(E)=\theta(\lambda)k_b \delta(E-E_b)+\frac{m}{\pi k}\frac{1}{1+\bk^2}
  \quad \to \quad \frac{1}{2} k_b \delta(E-E_b)
  + \frac{m}{\pi k}\left(\frac{1}{1+\bk^2}\right)_+, \label{eq:rhodistrtoy}
\end{equation}
where we have introduced the notation
\begin{equation}\label{eq:distrdef0}
  \left(\frac{1}{1+\bk^2}\right)_+=\sum_{i=0}^\infty \left([-\bk^2]^i\right)_+
\end{equation}
and where the $+$ on the terms of the expansion indicates that they are to be interpreted
as distributions defined by using an analytic regulator in $k$. In fact analytic
regularisation leads to
\begin{equation}
  \int_0^{E_{\rm cut}} m \frac{\mathd E}{k} E^n \left([-\bk^2]^i\right)_+
  = \theta_{n\ge i} \int_0^{k_{\rm cut}}\mathd k\, E^n(-\bk^2)^i - \theta_{n<i} \int_{k_{\rm cut}}^\infty \mathd k\,E^n(-\bk^2)^i.
  \label{eq:distrdef}
\end{equation}
Since $i$ is not greater than $n$ in the first integral,
and greater than $n$ in the second one, both integrals are convergent,
and do not require regularisation.

Integrating equation
(\ref{eq:rhodistrtoy}) in the energy up to $E_{\rm cut}$ and using
eqs.~(\ref{eq:distrdef}) and~(\ref{eq:distrdef0}), we recover eq.~(\ref{eq:In}).
We stress that also the first term of eq.~(\ref{eq:rhodistrtoy}) can be expanded
in powers of $\lambda$ with coefficients that are derivatives of $\delta(E)$,
to yield a full perturbative expansion for $\rho(E)$.

Finally, we stress again that the all-order expression~(\ref{eq:rhodistrtoy}) is
valid for both signs of $\lambda$, as one would expect from a perturbative expansion.

\subsection{The origin of the $1/2$ factor}\label{sec:onehalf}
We observe that in formula (\ref{eq:rhodistrtoy}) the substitution $\theta(\lambda) \to 1/2$
can also be justified heuristically as follows. The Taylor expansion in $\lambda$
of the bound state contribution involves only odd powers of $\lambda$ since $k_b=\lambda m$
and $E_b$ is quadratic in $\lambda$. The continuum contribution is instead a function of
$\lambda^2$. Its expansion may generate terms with odd powers of $\lambda$, but these terms
must depend upon $|\lambda|$ since the continuum contribution is even in $\lambda$.
The existence of a valid perturbative expansion for $\rho$ implies that by combining
the terms odd in $\lambda$, we must obtain contributions that do not involve
either theta functions or absolute values of $\lambda$. Thus, order by
order in perturbation theory, a relation of the form
\begin{equation}\label{eq:thetatotwo}
  A_j \lambda^j \theta(\lambda) + B_j |\lambda^j| = C_j \lambda^j,
\end{equation}
must hold,
where $A_j$ is the coefficient of the contribution of order $j$ from the bound state term,
$B_j$ is the one from the continuum term, and $C_j$ should be the one of their combination.
By imposing that this should hold for both $\lambda>0$ and $\lambda<0$ we get the
relation $C_j=A_j/2$, that justifies the substitution $\theta(\lambda)\to 1/2$.

\subsection{Direct perturbative calculation}
\label{sec:pertExp}
Since we know the full spectrum, we have been able to write down
the exact result for the resolvent, and then derive from it the full
perturbative expansion. By doing so one gets the impression that there
are terms in the perturbative expansion that arise from the combination
of bound-state effects, and of non-perturbative effects originating just
above the threshold for open top production, so that one wonders how
non-perturbative effects can give rise to perturbative ones.
In order to clarify this point, we can compute the resolvent directly using
perturbation theory. In the following we show how this is done by computing
the perturbative term of order $\lambda$.

We write
\begin{equation}
  R(E)=\frac{1}{H_0+V-E}=\frac{1}{H_0-E}-\frac{1}{H_0-E} V \frac{1}{H_0-E}+\ldots
\end{equation}
where
\begin{equation}
  H_0=-\frac{1}{2}\frac{\mathd^2}{\mathd x^2}, \quad\quad V=-\lambda \delta(x).
\end{equation}
The free eigenstates are
\begin{equation}\label{eq:freestates}
  \Psi_k^{(0)}(x) = \sqrt{\frac{2}{L}}\, \cos(k x).
\end{equation}
We neglect (as earlier) parity-odd solutions.
We obtain
\begin{eqnarray}
  R(E,0,0) &=&\sum_k \frac{|\psi_k^{(0)}(0)|^2}{E_k-E}+\lambda \sum_k\sum_{k'} 
               \frac{|\psi_k^{(0)}(0)|^2|\psi_{k'}^{(0)}(0)|^2}{(E_k-E)(E_{k'}-E)}\nonumber \\
  &=&
  \sum_k \frac{|\psi_k^{(0)}(0)|^2}{E_k-E}+\lambda \left(\sum_k
  \frac{|\psi_k^{(0)}(0)|^2}{(E_k-E)}\right)^2 \, .\label{eq:resfree}
\end{eqnarray}
We first compute the resolvent for $E<0$, where it is clearly analytic in $E$.
Using eq.~(\ref{eq:freestates}) and~(\ref{eq:resfree}),
and replacing $\sum_k=L/(2\pi)\int_0^\infty \mathd k$, we obtain
\begin{equation}
  R(E,0,0) = \sqrt{-\frac{m}{2E}} - \lambda \frac{m}{2E}\, .
\end{equation}
Its imaginary part above the real axis is
\begin{equation}
  R(E+i\epsilon,0,0) = i\theta(E)\sqrt{\frac{m}{2E}} + i\pi \frac{\lambda m}{2} \delta(E)
\end{equation}
leading to
\begin{equation}
  \rho(E) = \theta(E)\frac{m}{\pi k} +  \frac{\lambda m}{2} \delta(E)\,.
\end{equation}
This is equal to what we would obtain by expanding eq.~(\ref{eq:rhodistrtoy}) at order
$\lambda$ and using $k_b=\lambda m$.

Observe that in the calculation performed here no use has been
made of the existence of bound states, nor of the existence of effects
that arise from the full resummation of the perturbative expansion near
threshold. We find no trace of the funny partial cancellation of these two
effects, and of the ``magic'' factor of $1/2$. This is analogous
to what happens in QCD, where we know that the sum over all resonances
and open multiparticle states contributing to a spectral function
must match the result of a simple perturbative calculation. The only difference
in the present case is that we are able to compute the result in the
full theory. But, still, the result of the full theory
must match the one from perturbation theory, in ways that seem to be a
surprising combination of non-perturbative effects.

\section{The $t{\bar t}$ case}\label{sec:realcase}
We now consider the case of the $t\bar{t}$ system. It is equivalent to
the Hydrogen atom problem in quantum mechanics, except that the (reduced)
electron mass $\mu$ should be replaced by the top reduced mass
$\mu_t=m/2$, and that one should supply the colour factors
(\ref{eq:octectcoup}) and (\ref{eq:singletcoup}) to the coupling. The
spectral density has the form
\begin{eqnarray}
  \rho_l({\cal E}) &=& \theta(a_l) \times \frac{1}{\pi r_l^3} \sum_{n=1}^\infty \frac{1}{n^3} \delta({\cal E}-E_{l,n}) \label{eq:ttrho}
                +\frac{1}{4\pi^2}(m)^{3/2} \sqrt{{\cal E}} F(b_lv^{-1}) \\
  F(z)&=& \frac{z}{1-\exp(-z)},
\end{eqnarray}
where ${\cal E}=M_{t{\bar t}}-2m$, and $v=\sqrt{\cal E}/m$ is the velocity
of the quark in the $t{\bar t}$ rest frame.
The index $l$ is $1$ for the colour singlet and 8 for the colour octet state.
Furthermore
\begin{equation}
 r_l=\frac{2}{ma_l}, \quad E_{l,n}=-\frac{m}{4}\frac{a_l^2}{n^2},\quad b_l=\pi a_l,
\end{equation}
and 
\begin{equation}
  a_1=C_F\as, \quad a_8=-\frac{\as}{2N_C}.
\end{equation}
The factor $F(b_lv^{-1})$ is known as Sommerfeld factor~\cite{Sommerfeld,Sakharov}.

Notice that no bound states contributions exist for negative $a_l$ (i.e.
in the octet case).
We have defined a Bohr radius and eigenstate energies also
in this case, for future convenience.  The first term in
eq.~(\ref{eq:ttrho}) follows from the expression of the wave function
at the origin for zero angular momentum for the Hydrogen atom
\begin{equation}
  |\psi_{n00}(0)|^2=\frac{1}{\pi r^3 n^3},
\end{equation}
where $r$ is the Bohr radius $r=1/(\mu\alpha)$.  The normalisation of
the term in the continuum is fixed by the requirement that it should
match the free (i.e. $\as\to 0$) expression for zero coupling
\begin{equation}
  \sum_k\left|\frac{e^{i\vec{k}\cdot \vec{0}}}{\sqrt{V}}\right|^2 \delta\left(\frac{\vec{k}^2}{2\mu_t}-{\cal E}\right)
  =\frac{1}{(2\pi)^3}\int \mathd^3 k\delta\left(\frac{\vec{k}^2}{2\mu_t}-{\cal E}\right)= \frac{1}{4\pi^2}m^{3/2} \sqrt{{\cal E}},
\end{equation}
where $V$ is the (large) volume of the system.

Generalising what we found in eq.~(\ref{eq:rhodistrtoy}) of Sec.~\ref{sec:toymodel} we can immediately guess the formula for the
perturbative expansion of $\rho_l({\cal E})$ as
\begin{equation}\label{eq:rhoFull}
  \rho_l({\cal E}) \to   \frac{1}{2\pi r_l^3}\sum_{n=1}^\infty \frac{1}{n^3}\delta({\cal E}-E_{l,n})+\frac{m^{3/2}}{4\pi^2}\sqrt{\cal E}
  F^+(b_lv^{-1}),
\end{equation}
where $F^+$ stands for the Taylor expansion of the function $F$, with
coefficients interpreted as distributions to be regulated by the
analytic method. A detailed proof of eq.~(\ref{eq:rhoFull}) is given
in appendix~\ref{app:realcase}.

 As in the case of the toy model, in formula~(\ref{eq:rhoFull}) we have the
substitution $\theta(a_l)\to 1/2$, with $a_l$ playing the role of $\lambda$.
It is motivated by the same heuristic
argument given in the discussion around eq.~(\ref{eq:thetatotwo}). In the present
case, the continuum term is not fully even in $\lambda$, but due to the relation
\begin{equation}
  F(z) = \frac{F(z)+F(-z)}{2}+\frac{z}{2},
\end{equation}
stating that $F$ is the sum of an even function plus a linear term in
its argument, the same reasoning holds.

Using the expansion
\begin{equation}
  F(z)=1+\frac{z}{2}+\frac{z^2}{12}+{\cal O}(z^4)
\end{equation}
we find the expression for $\rho_l({\cal E})$ up to the first three orders in $\as$:
\begin{eqnarray}
  \rho_l({\cal E}) &=&  \frac{1}{2\pi r_l^3} \zeta(3) \delta({\cal E})+\frac{m^{3/2}}{4\pi^2}\sqrt{\cal E}
  \left(1+\frac{b_l}{2v} +\frac{b_l^2}{12v^2}\right) + {\cal O}(\as^4) \nonumber \\
            &=&\frac{m^2}{4\pi^2}\left(v+\frac{b_l}{2} +\frac{b_l^2}{12v}
                +\frac{\zeta(3)}{4\pi^2} b_l^3 m \delta({\cal E})
                + {\cal O}(\as^4) \right). \label{eq:rhottexpanded}
\end{eqnarray}
The first three terms, arising from the expansion of $F$,
give rise to convergent integrals, so
that the analytic regularisation is irrelevant for them.
We will see in the following that in practice no higher order terms are needed.
The cubic term in eq.~(\ref{eq:rhottexpanded})
was also computed in ref.~\cite{Beneke:2016jpx}, and we find full
agreement with that result.

\subsection{Summary of relevant results}
We now summarise our findings.  The inclusion of the leading threshold
contributions is achieved by replacing the factor $v$ (the top velocity in the
$t{\bar t}$ rest frame) in the Born
cross section (for each given colour configuration of the $t\bar{t}$
pair) with the factor
\begin{equation}
  v+\frac{b_l}{2}+\frac{b_l^2}{12v}+\frac{\zeta(3)}{4\pi^2}
  b_l^3  m\delta({\cal E}), \label{eq:enhancements}
\end{equation}
where ${\cal E}$ is the energy of the $t\bar{t}$ pair in its rest frame minus twice
the mass of the top. We remind here that the suffix $l$ is 1 for the singlet and
8 for the octet colour configuration of the $t\bar{t}$ pair, and
\begin{equation}
  b_1=\pi \Cf \as,\quad\quad b_8=\pi\left(-\frac{1}{2N_C}\right)\as.
\end{equation}

The corrections displayed in eq~(\ref{eq:enhancements}) enhance or
deplete the cross section near threshold. In the $gg$ fusion channel,
as discussed earlier, the spin singlet configuration is dominant at
threshold, and thus the corrections in eq~(\ref{eq:enhancements})
affect directly spin correlation observables.  They are formally of
NLO, NNLO and N$^3$LO order. Thus, if one is using a generator that is
NLO or NNLO accurate one should be careful to avoid over-counting.

\section{Modelling spin correlations near threshold}\label{sec:modeling}
We begin by studying the anatomy of spin correlations at the Born level.
It turns out that, at leading order, the top spin vector is given exactly
by the direction of the antilepton in the top rest frame. This fact
can be proven in two lines of spinor algebra, by writing the spin
structure of the top decay amplitude in term of the charge conjugate spinors
for the lepton, and then using a Fierz identity
\begin{equation}
  {\cal B}=\bar{u}_b \gamma^\mu(1-\gamma_5) u_t\;
  \bar{u}^c_\ell \gamma_\mu(1+\gamma_5)v^c_\nu 
  =\bar{u}^c_\ell(1-\gamma_5)u_t\;\bar{u}_b (1+\gamma_5) v^c_\nu,
\end{equation}
where the subscripts $b$, $\ell$ and $\nu$ refer to the $b$ quark, the lepton and the
corresponding spinor, and the superscript $c$ refers to charge conjugation. We can choose
the direction of the (anti)lepton in the top rest frame as spin quantisation
axis, i.e.
\begin{equation}
  s=\frac{m}{p_\ell\cdot p_t} p_\ell - \frac{1}{m}p_t
\end{equation}
which satisfies $s^2=-1$ and $s\cdot p_t=0$, and write the factor involving the top
quark spinor as
\begin{equation}
  \bar{u}^c_\ell(1-\gamma_5)u_t=\bar{u}^c_\ell(1-\gamma_5)\left(\frac{1+\gamma_5\slashed{s}}{2}
    +\frac{1-\gamma_5\slashed{s}}{2}\right)u_t.
\end{equation}
Using the Dirac equations for the spinors, we can easily verify that the second term
in the square bracket gives zero contribution, and thus that the top spin is a positive eigenstate of the component
of the spin operator in the direction of the antilepton. By CP invariance of the interaction leading to top decay,
we can immediately conclude that in the antitop case the top spin is the negative eigenstate of the
component of the spin operator in the direction of the lepton.

One defines the quantity
\begin{equation}
  \Chel{}=\hat{l}_1\cdot \hat{l}_2.
\end{equation}
The vectors $\hat{l}_1$  ($\hat{l}_2$) are defined as the direction of the lepton from top decay, in a frame obtained
by boosting the $t\bar{t}$ CM rest frame to the $t$ ($\bar{t}$) rest frame respectively.

By simple quantum mechanics, we can show that for a $t\bar{t}$ pair in the non-relativistic limit the angular distribution
of the leptons is such that
\begin{equation}
 \frac{1}{\sigma} \frac{\mathd \sigma}{\mathd \Chel{}}=\frac{1+\Chel{}}{2}
\end{equation}
Observe that anti-parallel $l_{1/2}$ means that the spins are align, which is incompatible with the singlet
configuration, and thus yields zero. We also obtain immediately
\begin{equation}
  \langle \Chel{} \rangle = \frac{1}{3},
\end{equation}
and one also introduces the quantity $D=-3 \langle \Chel{} \rangle$ that
equals $-1$ for a pure spin singlet state.

\subsection{The anatomy of spin correlations in $t\bar{t}$ production.}\label{sec:anatomy}
In this section we illustrate a simple, Leading-Order analysis of spin 
correlations in $t\bar{t}$ production and decay. As discussed earlier, we
expect that near threshold the $s$ wave production mechanism dominates, so that
the total angular momentum of the $t\bar{t}$ pair is given by its spin. The $gg$ fusion
production mechanism cannot yield a state of angular momentum 1, because of the Landau-Yang theorem,
so that in this case the $t{\bar t}$ system is in a spin singlet state. Conversely, in $q\bar{q}$ fusion
the $t{\bar t}$ must be in a spin triplet state, since the production goes through a spin 1 virtual gluon.

\begin{figure}[htb]
  \begin{center}
    \includegraphics[width=0.48\textwidth]{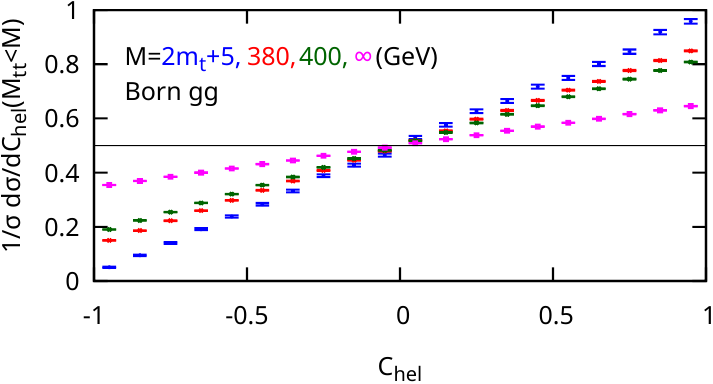}
    \includegraphics[width=0.48\textwidth]{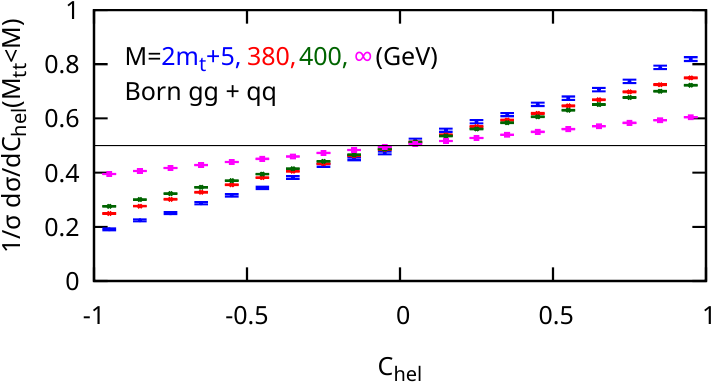}
  \end{center}
  \caption{\label{fig:bornchel} The distribution of the $\Chel{}$ observable
    computed at the Born level, by including only the $gg$ contribution (left plot), and by including
    all contributions (right plot). The distribution is presented for different upper limits on the
    invariant mass of the $t{\bar t}$ pair (where $M=\infty$ stands for no cut at all),
    with larger limits corresponding to smaller slopes.}
\end{figure}

We begin by illustrating in Fig.~\ref{fig:bornchel} the result for the $\Chel{}$ distribution at Born level.
We consider $pp$ collisions at $\sqrt{S}=13$~TeV with $m=172.5\;$GeV,
in the zero width limit, and for different cuts on the mass of the
$t{\bar t}$ pair (including the unrestricted case). We use the {\tt NNPDF30\_NLO\_as\_0118} pdf
set~\cite{NNPDF:2014otw}.\footnote{We could have used any other modern
  pdf set~\cite{Bailey:2020ooq,Hou:2019efy}, however the study of pdf
  sensitivity is outside the scope of the present work.}  It has been
carried out by generating events with the \POWHEGhvq{} generator,
setting to 1 the {\tt bornonly} and {\tt LOevents} flags in the {\tt
  powheg.input} file. Events are analysed at the Les Houches level.
As anticipated earlier, it is evident from the figure that the $gg$
channel contribution for the smallest mass cut $M=2m+5\,$GeV
approaches the $(1+\Chel{})/2$ distribution motivated by the Landau-Yang
theorem, as discussed earlier. It is also interesting to compare our
result for the unrestricted full distribution to the Standard Model
result of reference \cite{CMS:2024ynj}, displayed in the left plot of Fig.~2.
They seem in perfect agreement, showing that the elementary leading order
calculation presented in Fig.~\ref{fig:bornchel} already captures the
effects of a full simulation.

It is now instructive to implement the leading threshold corrections that we have computed in this work in the context of
the leading order $t{\bar t}$ production and decay.
The ${\cal O}(\as)$ and ${\cal O}(\as^2)$ corrections given in formula~(\ref{eq:enhancements}) can be straightforwardly
implemented as corrections to the Born process in the \POWHEGhvq{} process.
The third term, being concentrated at threshold,
requires some more attention. We have chosen to implement it by modelling the $\delta(E)$ function with a smooth function
concentrated near threshold, i.e. by replacing
\begin{equation}
  \delta(E)\approx\frac{105}{16E_c^{7/2}}\sqrt{E}(E_c-E)^2,
\end{equation}where $E=\sqrt{\mtt-2m}$, and we choose $E_c=1\;$GeV (we checked that
our results does not change when we double or halve $E_c$).

When implementing these corrections we have the freedom to choose whether to multiply them
by the exact Born level singlet and octet component of the cross section of equations
(\ref{eq:hqq}) and (\ref{eq:hgg}), or by just using their threshold limits.
We have chosen to multiply them by the full Born cross section, but perform
the separation of the octet and singlet channel according to the ratio $2/7$
for the singlet and $2/5$ for the octet, that is appropriate for the threshold
limit (corresponding to $\tau_1=\tau_2=1/2$ in formula (\ref{eq:hgg})).
We have verified that using the exact formulae for the ratio of the two
channels does not affect appreciably the result.

The choice of scales in the calculation requires particular attention. At the Born level, the renormalisation
scale used by \POWHEGhvq{} is taken by default equal to $\sqrt{m^2+p_\perp^2}$, where $p_\perp$
is the transverse momentum of the top quark. This scale is adequate to describe the top production process, and thus
is used in two powers of the coupling constant. The scale entering in the threshold enhanced corrections, should
instead be taken equal to the momentum of the top in the $t{\bar t}$ rest frame \emph{corresponding to the value of $M_{t{\bar t}}$ equal to the
  invariant mass cut.} Notice that, according to the discussion in the introduction of the present paper, the integral
of the cross section with a given invariant mass cut can be deformed into a contour integral that remains far from
the threshold region, so that it is the kinematic configuration at the cut that determines the scale.
Furthermore, a running scale choice taken as the top momentum on an event by event basis would lead
us to choose very low scale values, in particular for the term proportional to $\delta(E)$, in contrast with the
fact that an integrated measurement cannot depend upon the details of the
dynamics very near threshold.

In practice, we proceed as follows: we compute the distributions with the default coupling (as given by the \POWHEGhvq{})
generator, and add to the Born cross section the terms of eq.~(\ref{eq:enhancements}) computed with a fixed value
of $\as=\alpha_0$. We choose $\alpha_0=0.14$, but its value is irrelevant, as we will soon see.
We focus upon the calculation of a generic quantity $F(M)$, which depends upon a cut $M$ on the invariant mass of the $t{\bar t}$ pair.
Using the \POWHEG{}
reweighting feature, we are able to compute a Leading Order value, that we label $F^{(\rm LO)}(M)$,
and values that we label as $F^{(\le i)}_0$, that for $i=1$, 2 and 3 include the first, the first two, and all three
terms of eq.~(\ref{eq:enhancements}) evaluated with $\as=\alpha_0$. From these quantities we can reconstruct a cross
section where the scale entering the strong coupling in the threshold correction depends upon $M$, using the formula
\begin{eqnarray}
F^{(\le i)}_{\rm {\rm LO}}(M)  &=& F^{({\rm LO})}(M)+\theta(i\ge 1)(F^{(\le 1)}_0(M)-F^{({\rm LO})}(M))\frac{\as(S(M))}{\alpha_0} \nonumber \\
  &+&\theta(i\ge 2)(F^{(\le 2)}_0(M)-F^{(\le 1)}_0(M))\left(\frac{\as(S(M))}{\alpha_0}\right)^2 \nonumber \\
  &+&\theta(i\ge 3)(F^{(\le 3)}_0(M)-F^{(\le 2)}_0(M))\left(\frac{\as(S(M))}{\alpha_0}\right)^3 \label{eq:FiCorr}
\end{eqnarray}
where $F^{(\le i)}_{\rm {\rm LO}}$ represent the LO cross section corrected with threshold
enhanced contributions up to the order $i$.
$S(M)$ is our choice of the renormalisation scale as a function of $M$,
that we choose equal to one half of the top momentum in the $t\bar{t}$ rest frame:
\begin{equation} \label{eq:thrscale}
  S(M)=\frac{1}{4}\sqrt{M^2-4m^2}.
\end{equation}
The factor of one half is motivated by the fact that the distance of the
two quarks is twice the distance of each quark from the centre of the orbit,
that is conjugated to quark momentum.

When comparing to experimental results, suitable scale variations
will be introduced in order to estimate the uncertainty of our predictions, that
also account for the ambiguity in the choice of the Coulomb scale. More specifically, we consider
a standard seven-points scale variation, i.e. we vary the factorization and renormalisation scales by a factor
of two above and below their central value excluding only the cases when their ratio is four or one quarter,
 and define our error range as the region spanned
by the corresponding results. This is done in two alternative ways: in the first one we keep the renormalisation scale
in the threshold corrections at its central value; in the second one we multiply it by the
same renormalisation scale factor that we are using in the Monte Carlo calculation. We adopt as final
uncertainty range the union of the two ranges.

\begin{figure}[htb]
  \begin{center}
    \includegraphics[width=0.6\textwidth]{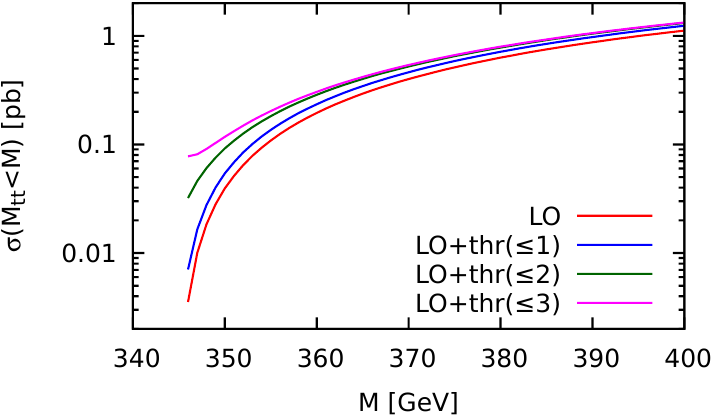}
  \end{center}
  \caption{\label{fig:mttcutplot-varalph} The $t\bar{t}$ cross section as a function
    of a cut on the invariant mass of the pair. Besides the Leading Order contribution, labelled LO,
    we plot the result obtained by including the leading threshold corrections up to the order
    ${\cal O}(\as)$, ${\cal O}(\as^2)$ and  ${\cal O}(\as^3)$, labelled as thr($\le$1), thr($\le$2) and thr($\le$3)
    respectively, that correspond to the expression $F^{(\le i)}_{\rm LO}$ of eq.~(\ref{eq:FiCorr}) for $i=1$, 2 and 3 respectively.
    The scale for the value of
    $\as$ associated with the threshold corrections is taken according to eq.~(\ref{eq:thrscale}).}
\end{figure}

We apply the procedure outlined above to the inclusive cross section as a function of an upper limit  $M$ on
the mass of the $t{\bar t}$ pair, and display the result in fig.~\ref{fig:mttcutplot-varalph}.

From the figure we see that threshold corrections become important for very low invariant mass cuts,
where they become comparable in size, while for large mass cuts they become smaller and decreasing with the order.

\begin{figure}[htb]
  \begin{center}
    \includegraphics[width=0.6\textwidth]{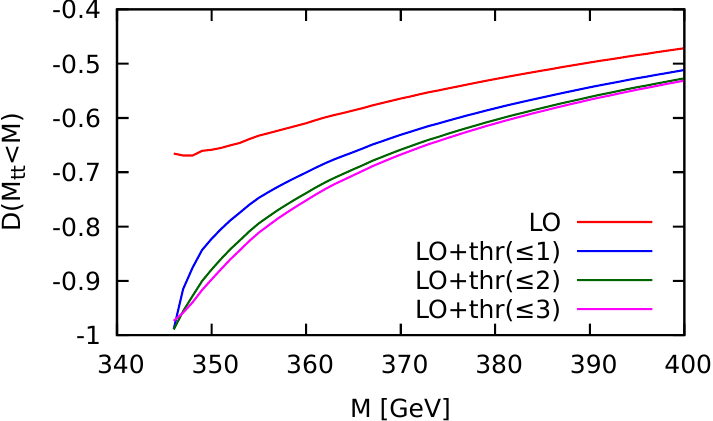}
  \end{center}
  \caption{\label{fig:dplot-varalph} The value of $D$ as a function
    of a cut on the invariant mass of the pair. Besides the Born contribution,
    we plot the result obtained by including the leading threshold corrections up to the order
    ${\cal O}(\as)$, ${\cal O}(\as^2)$ and  ${\cal O}(\as^3)$. The scale for the value of
    $\as$ are  defined as for fig.~\ref{fig:mttcutplot-varalph}. }
\end{figure}

In refs.~\cite{Beneke:2011mq,Beneke:2012wb,Beneke:2016jpx} the computation of the fully resummed total cross section for $t{\bar t}$ production is carried out, including the same threshold enhanced contributions of eq.~(\ref{eq:enhancements}) that we consider here. It is interesting to compare the contribution of the third-order term
found there. In table 1 of ref.~\cite{Beneke:2016jpx} (published version), the N$^3$LO threshold correction that we also compute is estimated
to be 0.63~pb for the fully inclusive cross section.
For the same correction we obtain 0.78~pb for a mass cut of 600~GeV. Since this
contribution is located very near the production threshold, its only dependence through the mass cut is due to the
argument of $\as$, that is set according to eq.~(\ref{eq:thrscale}), i.e. equal to 123~GeV.
In view of the
fact that ref.~\cite{Beneke:2016jpx} uses a slightly larger top mass and different PDF's, that we have a cut on the
invariant mass (which \emph{increases} the result, due to the corresponding larger value of $\as$),
and that the number is subject to large scale uncertainties, we consider this an acceptable difference. 
In Fig.~\ref{fig:dplot-varalph} we show the value of $D$ as a function of a cut in the invariant
mass of the $t{\bar t}$ system, i.e. the quantity
\begin{equation}
  D(\mtt<M)=\frac{\int\theta(M-\mtt) \mathd \sigma (-3 \Chel{})}
  {\int\theta(M-\mtt) \mathd \sigma}.
\end{equation}
The same criteria for the choice of the scale in $\as$ that have been used for fig.~\ref{fig:mttcutplot-varalph}
are also used in this case, where the numerator and denominator are computed with the same choice of scales.
The typical mass cuts $M$ used by the experimental analysis of this observable are around $380-400$~GeV.
In this region the NLO ($1/v$) corrections are quite important, the NNLO ($1/v^2$) corrections are non-negligible,
while the N$^3$LO ($\delta(E)$) corrections are very small. The erratic behaviour below $350$~GeV reminds
us that our calculation becomes meaningless very close to threshold.

\begin{figure}[htb]
  \begin{center}
    \includegraphics[width=0.6\textwidth]{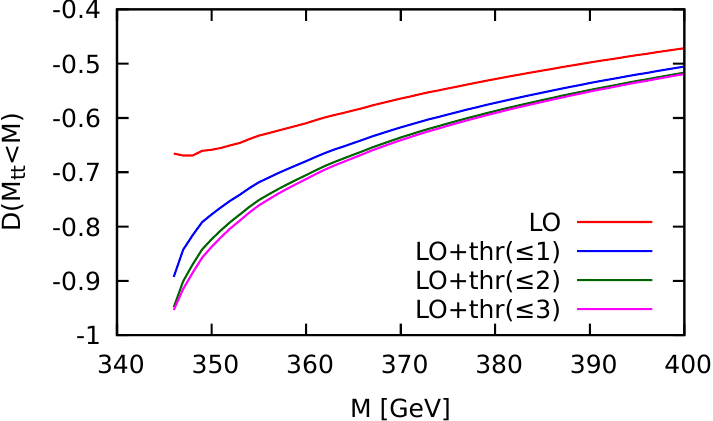}
  \end{center}
  \caption{\label{fig:dplot-constalph} As in fig.~\ref{fig:dplot-varalph}, but with
  $\as$ evaluated at the \POWHEGhvq{} default scale also for the threshold contributions.}
\end{figure}

In order to verify the importance of the
correct scale choice we also show in Fig.~\ref{fig:dplot-constalph}
the result one would obtained by using the \POWHEGhvq{} default scale
also for the threshold contributions. It is apparent from the figure
that there is a non-negligible difference with respect to the result obtained with the
appropriate scales in the threshold correction factors. 
Typically, NLO+PS Monte Carlo generators implement the radiative corrections using scales that are associated with
the hard production process, and do not distinguish effects that are dominated by very different scales.
When computing threshold effects to correct NLO+PS and also NNLO+PS prediction we will properly account
for the effect of the appropriate scale choice in the threshold correction terms.

\subsection{Nominal Monte Carlo Results at the NLO+PS level}\label{sec:nominal}
In this section we compare the Monte Carlo results for the invariant mass of the top pair and for the $D$ parameter
as a function of an invariant mass cut. We stress that, at the NLO+PS level, the first term in eq.~(\ref{eq:enhancements})
is already included in the NLO calculation, but it is evaluated at the standard NLO+PS scale. We will discuss later
the effect due to the use of a more appropriate scale for this term. In the present section we only consider the
MC predictions obtained with nominal Monte Carlo implementations.

Our default setup is for $pp$ collisions at $\sqrt{S}=13$~TeV with
$m=172.5\;$GeV, finite top width equal to 1.31~GeV, and using the {\tt
  NNPDF30\_NLO\_as\_0118} pdf set.  We only present results obtained
at the Les Houches parton level. We have verified that considering
parton level results after showering with
\Pythiaeight~\cite{Bierlich:2022pfr} does not lead to appreciable changes
for the observables we are interested in. Furthermore, in the present
work we are interested into assessing the relative effects of
threshold corrections, and will not consider possible shower
dependencies.

\begin{figure}[htb]
  \begin{center}
    \includegraphics[width=0.6\textwidth]{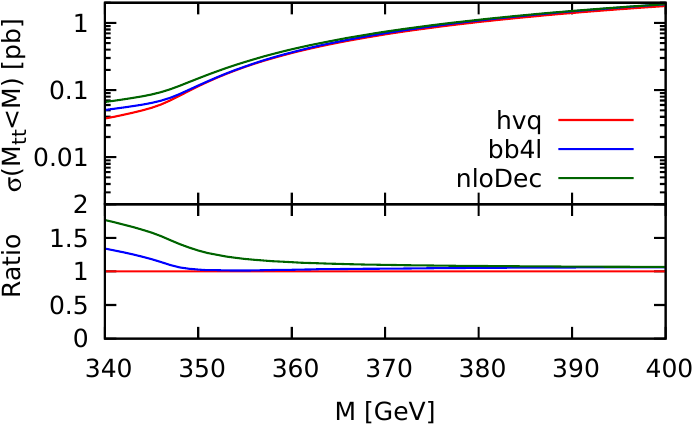}
  \end{center}
  \caption{\label{fig:mttcutplot-MC} Integrated cross section for $t{\bar t}$ production as
    a function of a cut on the pair mass.}
\end{figure}

\begin{figure}[htb]
  \begin{center}
    \includegraphics[width=0.6\textwidth]{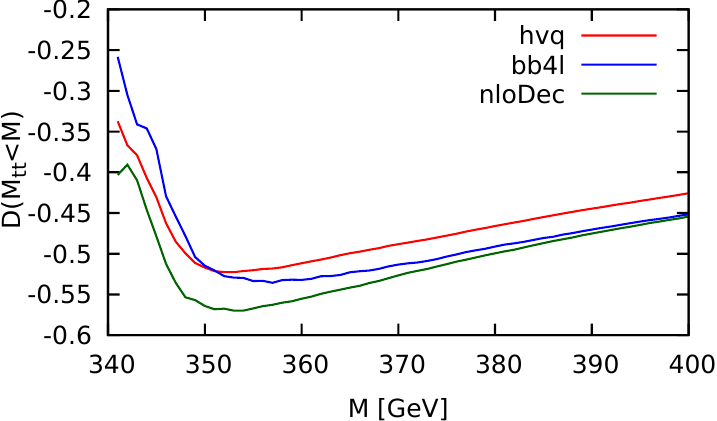}
  \end{center}
  \caption{\label{fig:dplot-MC} The $D$ value as a function of the invariant mass cut
    for our three generators of choice.}
\end{figure}

We begin by showing in fig.~\ref{fig:mttcutplot-MC} and \ref{fig:dplot-MC}
the results obtained by using the \POWHEGhvq{} \cite{Frixione:2007nw}, the \POWHEGnloDec{} \cite{Campbell:2014kua},
and the \POWHEGbbfourl{} \cite{Jezo:2016ujg} generators, and
show also their ratio relative to \POWHEGhvq{}. We notice that there are differences among the different codes,
especially when approaching the threshold region. Furthermore, for the $D$ observable sizeable differences
remain also at relatively large invariant mass cuts.

\begin{figure}[htb]
  \begin{center}
    \includegraphics[width=0.6\textwidth]{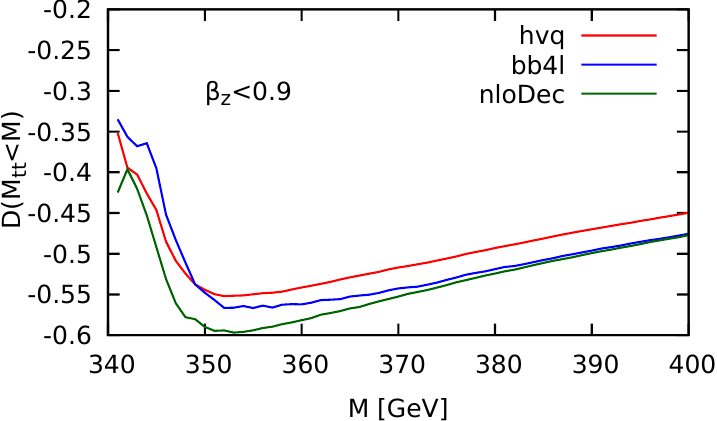}
  \end{center}
  \caption{\label{fig:dplot-MC-betaz} The $D$ value as a function of the invariant mass cut
    for our three generators of choice, with the restriction $\beta_z<0.9$ on the longitudinal
    velocity of the $t{\bar t}$ system.}
\end{figure}

In ref.~\cite{CMS:2024pts} the $D$ observable is measured with an auxiliary restriction on the longitudinal
velocity of the $t{\bar t}$ system $|\beta_z|<0.9$, in order to enhance the correlations by reducing the
contribution from the $q{\bar q}$ channel.  In fig.~\ref{fig:dplot-MC-betaz}
we present our result when implementing this cut.
It is interesting to compare the value of $D$ for $M=400$~GeV in fig.~\ref{fig:dplot-MC-betaz} to the one
in fig.~9 of ref.~\cite{CMS:2024pts}. The number
corresponding to the \POWHEGhvq{} prediction in figure~\ref{fig:dplot-MC-betaz},  roughly equal to $D=-0.45$,
compares well with the result reported in the upper part of the CMS figure
(the orange band with the dashed line).\footnote{From the hepdata database, the corresponding number is $-0.452 \pm 0.0017\pm 0.0037$}

The differences in the results obtained with the different Monte Carlo
generators studied in this section seem to be due only to the
accuracy with which the top decay have been implemented. The
\POWHEGhvq{} code uses the approximate algorithm proposed in
ref.~\cite{Frixione:2007zp}. A general
automatic implementation of the algorithm of ref.~\cite{Frixione:2007zp}, dubbed MadSpin, was proposed
in ref.~\cite{Artoisenet:2012st}, but~\POWHEGhvq{} uses its own implementation.
Both \POWHEGnloDec{} and \POWHEGbbfourl{} implement an increasingly more accurate
description of the decay process. From our figure, however, it is hard to judge
if the two more advanced codes give consistent results. We have also interfaced
\POWHEGhvq{} with the MadSpin code, and found a result (not shown in the figures)
consistent with the \POWHEGnloDec{} when $M>360$~GeV, but departing rapidly from it
for smaller $M$. Overall, these comparisons seem to suggest that the \POWHEGnloDec{},
\POWHEGbbfourl{} and \POWHEGhvq{} with MadSpin results should be preferred over the
\POWHEGhvq{} one for the observables we are considering. We believe, however, that in order to draw a firm conclusion
a dedicated Monte-Carlo study should be performed, analysing more observables than
the ones used in the present context in order to understand the origin of the differences.

\subsection{Inclusion of the enhanced threshold corrections in fixed order calculation}\label{sec:nloimproved}
The inclusion of the enhanced threshold corrections when starting with a NLO
calculation can be performed according to the same line of reasoning followed when
starting with the Born cross section. Care must be taken, however, to subtract the $\as/v$
term that is already included in the NLO calculation, in order to avoid double counting.
We adopt the formula
\begin{eqnarray}
F^{(\le i)}_{\rm NLO}(M)  &=& F^{(\rm NLO)}(M)+\theta(i\ge 1)(F^{(\le 1)}_0(M)-F^{(\rm LO)}(M))\frac{\as(S(M))-\as(S'(M))}{\alpha_0} \nonumber \\
  &+&\theta(i\ge 2)(F^{(\le 2)}_0(M)-F^{(\le 1)}_0(M))\left(\frac{\as(S(M))}{\alpha_0}\right)^2 \nonumber \\
  &+&\theta(i\ge 3)(F^{(\le 3)}_0(M)-F^{(\le 2)}_0(M))\left(\frac{\as(S(M))}{\alpha_0}\right)^3
\end{eqnarray}
where, as before, $F^{(\le i)}_0$ represents a LO result corrected by adding the threshold enhanced
corrections up to the order $i$, and evaluated with the coupling $\alpha_0$.
$S'(M)$ represents the average value of the renormalisation scale used in the NLO calculation
for the observable of interest. In our specific case, it is given by the average value
of $\sqrt{p_\perp^2+m^2}$ as a function of the invariant mass cut $M$. We can easily compute $S'(M)$
using our generator. Notice that  $S'(M)$ remains large (i.e. greater than $m$) even close to the production
threshold, which is a necessary condition for this procedure to be applicable.

\begin{figure}[htb]
  \begin{center}
    \includegraphics[width=0.6\textwidth]{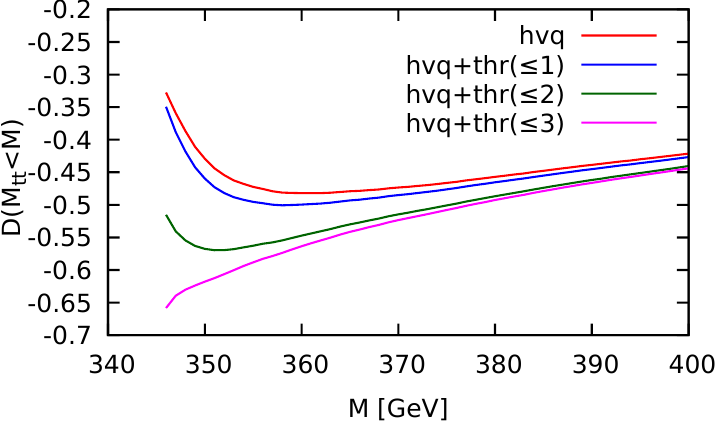}
  \end{center}
  \caption{\label{fig:dplot-NLO} The $D$ value as a function of the invariant mass cut
    for the \POWHEGhvq{} generator, when the result is amended with the
    inclusion of the dominant threshold corrections up to the N$^3$LO order.}
\end{figure}

\begin{figure}[htb]
  \begin{center}
    \includegraphics[width=0.6\textwidth]{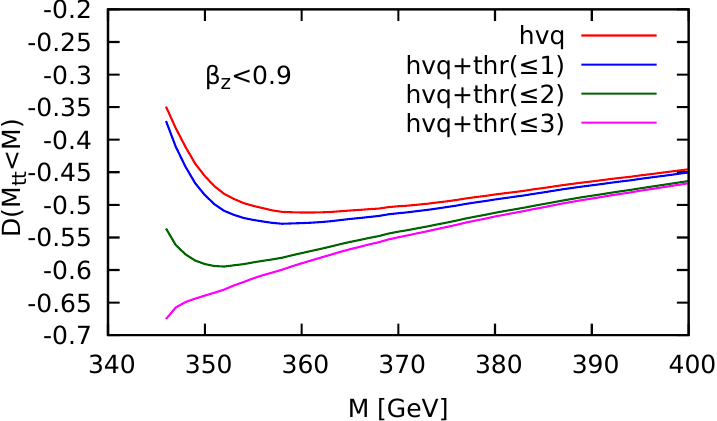}
  \end{center}
  \caption{\label{fig:dplot-NLO-betaz} As in fig.~\ref{fig:dplot-NLO} when the $\beta_z<0.9$ cut is applied.}
\end{figure}

\begin{figure}[htb]
  \begin{center}
    \includegraphics[width=0.6\textwidth]{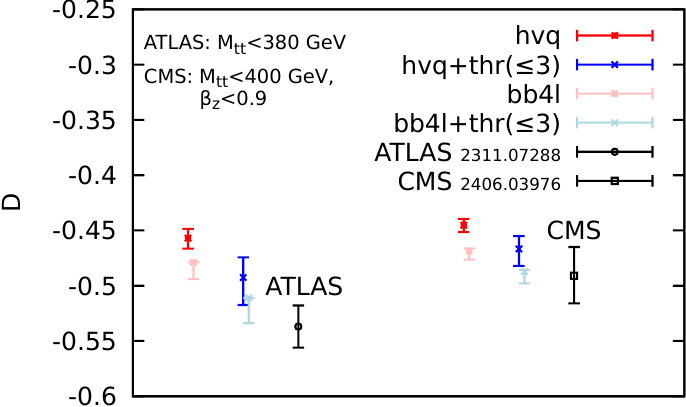}
  \end{center}
  \caption{\label{fig:Prediction} Prediction for the $D$
    observable as measured by ATLAS and CMS, obtained
    with the nominal \POWHEGhvq{} and \POWHEGbbfourl{} generators,
    and with the inclusion of the enhanced threshold corrections
    up to N$^3$LO order.}
\end{figure}

Our results for the $D$ function obtained with the method outlined above
are displayed in Figs.~\ref{fig:dplot-NLO} and \ref{fig:dplot-NLO-betaz}
for the cases with or without the cut on the longitudinal velocity of the $t\bar{t}$ system.
Furthermore, in table~\ref{tab:DAtlasCms} we show our prediction for the
$D$ measurements from the ATLAS~\cite{ATLAS:2023fsd} and CMS~\cite{CMS:2024pts} collaborations.
\newcommand\mystack[2]{\begin{smallmatrix} #1 \\ #2 \end{smallmatrix}}
\begin{table}[htb]
{\footnotesize
  \begin{center}
    \begin{tabular}{|l|l|c|c|c|c|c|}
      \hline
      Exp. & MC & nominal & +thr($\le$1) & +thr($\le$2) & +thr($\le$3) & data \\
      \hline
      ATLAS & hvq & $-0.457\mystack{ +0.008}{-0.010}$  & $-0.465\mystack{ +0.011}{-0.013}$  & $-0.487\mystack{ +0.016}{-0.022}$  & $-0.493\mystack{ +0.018}{-0.025}$ & $-0.537 \pm 0.019$  \\
      \cline{2-6}
           & bb4l & $-0.479\mystack{ +0.000}{-0.015}$  & $-0.486\mystack{ +0.000}{-0.015}$  & $-0.506\mystack{ +0.000}{-0.021}$  & $-0.511\mystack{ +0.000}{-0.023}$  &  \\
      \hline 
      CMS  & hvq & $-0.445\mystack{ +0.006}{-0.006}$  & $-0.450\mystack{ +0.007}{-0.008}$  & $-0.463\mystack{ +0.011}{-0.014}$  & $-0.467\mystack{ +0.012}{-0.015}$ & $-0.491\mystack{+ 0.026}{ - 0.025}$ \\ 
      \cline{2-6}
           & bb4l & $-0.468\mystack{ +0.002}{-0.008}$  & $-0.473\mystack{ +0.004}{-0.007}$  & $-0.484\mystack{ +0.002}{-0.009}$  & $-0.487\mystack{ +0.001}{-0.011}$ &  \\
      \hline
    \end{tabular}
  \end{center}
}
\caption{\label{tab:DAtlasCms} Predictions compared with data for the
$D$ measurements from the ATLAS~\cite{ATLAS:2023fsd} and CMS~\cite{CMS:2024pts} collaborations.}
\end{table}
In order to better appreciate visually the status of the comparison, we also display it in fig~\ref{fig:Prediction},
where we also include results obtained with the \POWHEGbbfourl{} generator.
As can be seen from the table and the figure, no relevant differences between the data and the
theoretical predictions including threshold effects can be seen. The observation remains,
however, that the \POWHEGhvq{} implementation yields slightly less correlations then the
\POWHEGbbfourl{} one, consistently with what we found in Section~\ref{sec:nominal}

\subsection{Results for a NNLO generator}\label{sec:minnlo}
We now describe the implementation of our result in the NNLO generator \ttMiNNLO{}. In this case,
threshold enhanced contributions are already included both at NLO and NNLO. Furthermore, besides the
correction of relative order $\as/v$, also those of order $\as^2/v$ are included. This implies that
some sort of scale compensation is taking place, so that we cannot add the
difference of the $\as/v$ contributions evaluated at two different scales, as we did
in the NLO case. It makes sense, however,
to add the $(\as/v)^2$ correction with $\as$ evaluated at the scale of the threshold cut, and subtract the
same quantity with $\as$ evaluated at the hard scale of the process. The delta term, of order
$\as^3$, can be added as before.

\begin{figure}[htb]
  \begin{center}
    \includegraphics[width=0.6\textwidth]{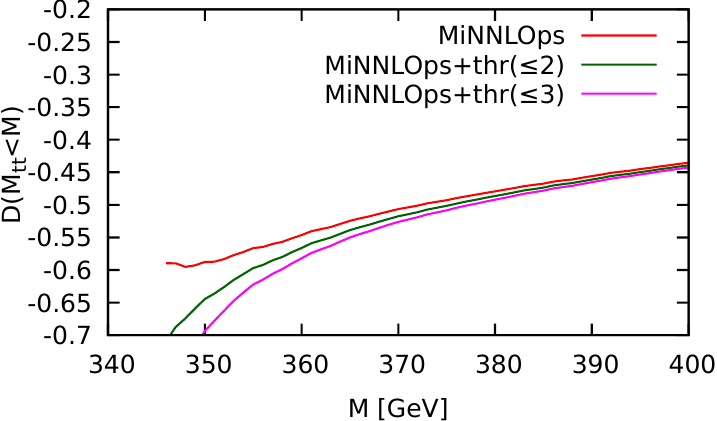}
  \end{center}
  \caption{\label{fig:dplot-NNLO} Prediction for the $D$
    observable, obtained
    with \ttMiNNLO{},
    and with the inclusion of the enhanced threshold corrections
    up to N$^3$LO order.}
\end{figure}

\begin{figure}[htb]
  \begin{center}
    \includegraphics[width=0.6\textwidth]{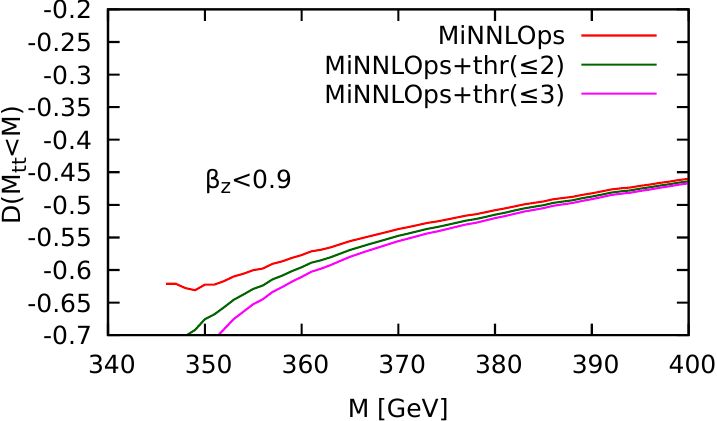}
  \end{center}
  \caption{\label{fig:dplot-NNLO-betaz} As in fig.~\ref{fig:dplot-NNLO} in presence of the
    cut $\beta_z<0.9$.}
\end{figure}

\begin{figure}[htb]
  \begin{center}
    \includegraphics[width=0.6\textwidth]{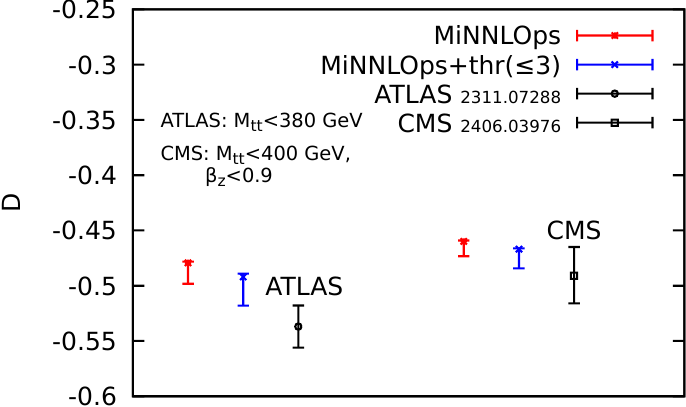}
  \end{center}
  \caption{\label{fig:PredictionNNLO} Prediction for the $D$
    observable as measured by ATLAS and CMS, obtained
    with the nominal \ttMiNNLO{} generators,
    and with the inclusion of the enhanced threshold corrections
    up to N$^3$LO order.}
\end{figure}

We begin by showing in figs.~\ref{fig:dplot-NNLO}
and ~\ref{fig:dplot-NNLO-betaz}
the predictions for the $D$ observable obtained using the nominal \ttMiNNLO{} generator, the prediction
obtained by adding the difference of the $\as^2/v^2$ term evaluated at the Coulomb scale minus
the same term evaluated at the hard scale, and the prediction obtained by also adding the $\as^3$
delta term. We see that around the relevant scales of $380$-$400$~GeV these corrections are much smaller
than the corrections found when using an NLO generator. The relevant numbers
for the ATLAS and CMS experimental configurations are given in table~\ref{tab:MiNNLOpred}, and
are illustrated in Fig.~\ref{fig:PredictionNNLO}.
\begin{table}[htb]
  {\footnotesize
  \begin{center}
    \begin{tabular}{|l|c|c|c|c|}
      \hline
      Exp. & MiNNLO & MiNNLO+thr($\le$2) & MiNNLO+thr($\le$3) & data \\
      \hline
      ATLAS & $-0.479\mystack{ +0.001}{-0.019}$ & $-0.486\mystack{ +0.001}{-0.023}$  & $-0.492\mystack{ +0.003}{-0.026}$  & $-0.537 \pm 0.019$ \\
      \hline
      CMS   & $-0.460\mystack{ +0.001}{-0.013}$  & $-0.464\mystack{ +0.001}{-0.015}$ & $-0.467\mystack{ +0.001}{-0.017}$ & $-0.491\mystack{+ 0.026}{ - 0.025}$   \\
      \hline
    \end{tabular}
  \end{center}
}
\caption{\label{tab:MiNNLOpred} Prediction for the $D$ observable using \ttMiNNLO{} in the cases corresponding
  to the ATLAS and CMS cuts, compared with data.}
\end{table}
As for the NLO case, we observe good agreement between the prediction and the experimental
results, where now, however, also the nominal Monte Carlo yields fairly good agreement with
data. Unlike the NLO case, we are unable to present a study of the impact of the correlation
model. The \ttMiNNLO{} generator uses the same method of the \POWHEGhvq{} generator for
the implementation of spin correlations, and we do not have alternative implementations
of higher accuracy. We observe, however, that fixed-order NNLO calculation of top production
and decay do exist in the literature~\cite{Gao:2017goi,Behring:2019iiv,Czakon:2020qbd},
and may be used to assess these effects (possibly by supplementing them with the NLO electroweak corrections for
off-shell top pair production~\cite{Denner:2016jyo}).

\begin{figure}[htb]
  \begin{center}
    \includegraphics[width=0.6\textwidth]{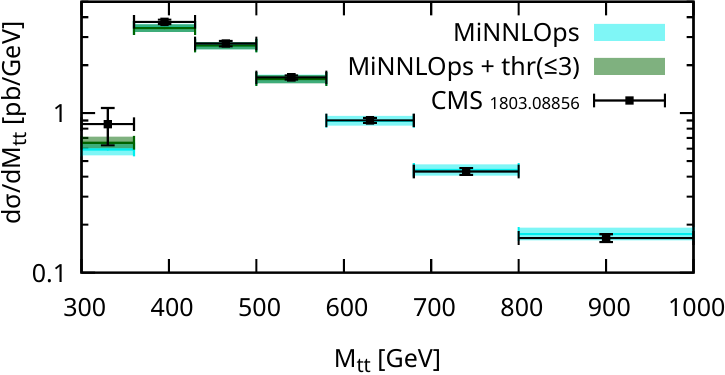}
  \end{center}
  \caption{\label{fig:CMSold} Invariant mass distribution
    of the $t{\bar t}$ pair compared to CMS data. The MiNNLO predictions
    are shown with and without the inclusion of threshold corrections
    up to the cubic term.}
\end{figure}

As a last point, we would like to consider the prediction for the
differential distribution in the $t{\bar t}$ mass obtained in the
original \ttMiNNLO{} publication \cite{Mazzitelli:2021mmm}, and display in Fig.~2 of that work,
where the MiNNLO prediction showed a visible deficit in the first bin
as compared to the CMS measurement of ref.~\cite{CMS:2018htd}. We are
now in the position to add the threshold corrections to the MiNNLO
prediction, and display the result in Fig.~\ref{fig:CMSold}.
We can see that with the addition of the threshold corrections,
although going in the right direction, is not quite enough the resolve
the discrepancy.  The relevant numbers for the results in the first
bin are $0.592+0.058- 0.048$ for the pure MiNNLO result, while for the
fully corrected result (i.e thr($\le 3$)) we get
$0.652+0.062-0.053$. When including only threshold corrections up to
the second order (i.e thr($\le 2$)) we get instead the result
$0.626+0.055- 0.049$, indicating that the ${\cal O}(\as^3)$ threshold
correction is slightly less than the correction to the
${\cal O}(\as^2)$ contribution arising from the mismatch of the scale
choice in the \ttMiNNLO{} generator and in our calculation of the
threshold terms. In summary, the corrections that we compute go in the
right direction but remain however smaller than the CMS measurement
that equals $0.853 \pm 0.225$, obtained by rescaling the CMS (normalised) result
by the total cross section of $832\;$pb.

\section{Comparison with alternative calculations of threshold effects.}
The formalism to deal with non-relativistic effects in $t\bar{t}$ production has
been available for a long time~\cite{Fadin:1987wz,Fadin:1988fn,Fadin:1990wx}, both for $e^+e^-$
and for hadron colliders, and more applications relevant to hadron colliders have appeared
in ref.~\cite{Sumino:1992ai,Hagiwara:2008df,Kiyo:2008bv,Sumino:2010bv}. These approaches allow for a detailed
description of the $m_{t{\bar t}}$ spectrum near the threshold region, by combining non-relativistic
and finite width effects. The standard measurements of $m_{t{\bar t}}$ at hadron colliders cannot
resolve the detailed of the spectrum computed in this framework.\footnote{There is however a context
where such precision may be reached, i.e. in the measurement of the contribution of top quark loops
to diphoton production~\cite{Kawabata:2016aya,Dugad:2016kdv}.}

As pointed out in this work, whenever the experimental resolution of $m_{t{\bar t}}$ is large (i.e. of the order of tens of GeV)
one can obtain the non-relativistic effects as a simple power expansion in $\as/v$, where $v$ is the typical
velocity of the top in the $t{\bar t}$ rest frame that is allowed by the resolution on the invariant mass. This perturbative expansion
is easily matched with fixed order calculation. We have shown that $\as/v$ and $(\as/v)^2$ are the dominant effects,
while the contribution of order $\as^3$ is localised very near threshold, and it comprises bound state effects, but also
effects above threshold that enter with similar magnitude.

We observe that in the relevant literature on threshold effects much emphasis is put on the formation of bound states. In reality
the formalism of refs.~\cite{Fadin:1987wz,Fadin:1988fn,Fadin:1990wx} also includes effects of order $\as$ and $\as^2$ that
cannot arise from bound states. The question now arises about to what extent the two formulations are equivalent. As we
have seen in this work, if we consider only the leading non-relativistic effects, which correspond to
a full resummation of the $(\as/v)^n$ enhanced contributions, our formulation is equivalent to a perturbative
expansion of the fully resummed result. On the other hand, it is also clear that our approach implies differences
arising from the running of the strong coupling constant that are not easily incorporated in the fully resummed result.
We thus find that there is an intrinsic difficulty in modelling these corrections in a Monte Carlo generator framework.
Monte Carlos are supposed to yield a good description of the phenomena at all scales involved. On the other hand,
if we want that our Monte Carlos' description accurately describes the shape of the $m_{t{\bar t}}$ distribution near threshold,
they should generate the configurations near threshold with a strong coupling constant evaluated at a scale corresponding to
the solution of the self-consistency equation $\as(m_t v)/v \approx 1$. This results in a value of $\as$ that is inappropriate
when considering inclusive measurements covering a wide range of the invariant mass, where, as we have seen, the scale
should be taken from formula (\ref{eq:thrscale}), leading to appreciably smaller values of $\as$. Ideally, the Monte Carlo
should yield large corrections in the region near threshold, compensated by negative corrections spread out above threshold
in order to reproduce the correct inclusive results. In principle, calculations using non-relativistic QCD beyond the leading
order~\cite{Garzelli:2024uhe} may be able to do this, but their implementation in a Monte Carlo generator may turn out to be not so
straightforward.

As far as comparison of the theoretical predictions that we have obtained with experimental measurements,
it should not be too difficult to develop an approximate implementation
of a Monte Carlo generator that provides an adequate description of the data.
However, it is desirable that the experimental results on threshold cross
section excesses be made available as unfolded cross sections in the near future,
as this would allow for a more reliable comparison with theoretical predictions.

\section{Conclusions}\label{sec:conclusions}
Spin correlations in top pair production are strongly enhanced near the threshold region, which
suggests that phenomena induced by Coulombic strong interactions in the non-relativistic regime
should affect them in an important way. This fact has led several authors to consider the inclusion
of $t{\bar t}$ bound state modelling, in order to explain some
difficulties that one encounters when attempting to describe $t{\bar t}$ spin correlations data using standard Monte Carlo models.

In the language of QCD perturbation theory, as one approaches the threshold for top pair production (i.e. as the mass of the
$t{\bar t}$ pair $M_{t{\bar t}}$ approaches twice the mass of the top $m$) corrections that scale as $(\as/v)^n$, where $v$ is the velocity
of the top quarks in the $t{\bar t}$ rest frame and $\as$ is evaluated at the scale $m v$, arise at all orders in perturbation
theory. Eventually, when $v\approx \as$, these corrections become of order 1, and one should resum all of them in
order to get a sensible result. The ensuing physics, if the Coulomb interaction is
attractive, also involves the formation of bound states. In full analogy with the Hydrogen atom, bound states are characterised by
$v \approx \as$, momenta of order $m\as$, a size of order $1/(m\as)$, and energies of order $m\as^2$. The time it takes for a quark with velocity $\as$
to cover a distance of order $1/(m\as)$ is $1/(m \as^2)$. This is of the order of few GeV, a time scale that is comparable
with the top lifetime. For this reason the formation of a narrow bound state is not possible. Yet, ``lucky'' top quarks that
live longer than the top (average) lifetime may be able to circle the orbit a few times, so that the bound state
can in fact manifest itself as a small bump in the production cross section. In order to compute the detailed structure
of this feature one needs to resum the perturbative expansion to all orders in the enhanced  $(\as/v)^n$ corrections, which
amounts to solving the non-relativistic bound state equation.

In the present work we have shown that it is not necessary to fully solve the bound state problem in order to
compute spin correlations effects in $t{\bar t}$ production and decay. The key observation is that the experimental resolution in the
$t{\bar t}$ mass measurement is quite large, and in fact the observables used by the experimental collaborations for
spin correlation measurements
are integrated cross sections up to a relatively large mass cut $M\approx 400$~GeV. We have demonstrated that, this being the
case, the enhanced contributions to these cross sections scale as $(\as/v_{\rm cut})^n$, where $v_{\rm cut}$ is the
velocity of the top quark in the $t{\bar t}$ rest frame when the mass of the $t{\bar t}$ system is of order $M$, that yields
$v_{\rm cut}\approx 0.29$. Furthermore, the scale at which $\as$ should be evaluated to compute these contributions is the
corresponding momentum of the top quark, that is around $50$~GeV, well in the perturbative regime but considerably smaller
than the hard production scale, that is near the top mass. We have computed the coefficients of the threshold enhanced terms
in the perturbative expansion to all orders. We find that the first three terms, of relative order up to $\as^3$,
yield already a sufficient precision, the third term giving a tiny contribution. The first two terms in the expansion are well known, while the third order term
was already computed by Beneke and Ruiz-Femenia~\cite{Beneke:2016jpx} for fully inclusive cross sections,
and we agree with their result.

Since we are able to write a formula for the fully resummed result, we can trace the various contributions to the terms
up to the third order. The first and second order terms are well-known Coulomb enhanced contributions of relative
order $\as/v$ and $(\as/v)^2$, not related to the formation of bound states. Bound states, which are present in the
colour singlet channel of the production cross section, start contributing at order~$\as^3$ (see section~\ref{sec:powercount}).
However, there are also contributions from open top production that have exactly the same form, and reduce the
bound states contribution by exactly a factor of 2. This feature is present in both our toy
model and in the case of the Coulomb potential,
and is in part due to the fact that the bound state contribution only exists for positive coupling,
and thus, if it was the only contribution of that form, would not be compatible with a perturbative expansion.

In view of the inclusive character of the corrections that we compute, we cannot implement them directly into
Monte Carlo generators. Instead, we include in the generators the contribution of this effects evaluated
with a fixed reference coupling constant. Then we compute the relevant distributions, that are always
function of a mass cut $M$, and rescale the contributions of the threshold effect to the value of $\as$
evaluated at a scale of the order of the top momentum corresponding to $M$. When using our
threshold corrections to augment results obtained with NLO or NNLO generators, we take care to subtract
the threshold contributions evaluated at the hard scale of the process, which are already included in the
generators.

Regarding the contribution of the various terms, we find that the most important ones are those
of relative order $\as/v$ and $(\as/v)^2$. Furthermore, when using the NLO generators, we find
a non-negligible contribution of the  $\as/v$ term due to the change in the scale of the coupling constant that
we operate, a sizeable term of order $(\as/v)^2$, and a smaller correction from the $\as^3 \delta(M_{t{\bar t}}-2m)$
term.

When comparing with published data, we find that the corrections that we compute reduce considerably the tension
between theoretical predictions and data. We also found, however, that there is another effect, due to the
approximate treatment of top decays in \POWHEGhvq{}, that may contribute to the tension, since
generators that implement more accurately the top decay process lead to slightly stronger correlations.
Further work in the framework of NLO+PS and NNLO+PS may be needed to fully clarify this point.

Several attempts have appeared in the literature to explicitly add a model for the formation of a $t{\bar t}$ bound
state to compensate for missing threshold contributions in the description of spin correlations in $t{\bar t}$
production and decay. The effect of such addition would lead to corrections that do not widely differ from
the one arising from the three threshold terms that we consider in the present work, since they all model
$s$-wave corrections that are enhanced in some way near threshold. When turning these models into solid
QCD predictions, however, some care must be taken. First of all, it is not enough to consider bound
states production, since there are continuum contributions that have the same form and similar size.
In contrast, a full treatment of threshold effects using non-relativistic QCD would of course include all necessary
corrections. However the accuracy of such treatment should be such that the integral of the production
cross section up to relatively large cut in the invariant mass of the $t{\bar t}$ pair should satisfy
the perturbative expansion that we found. In other words, the perturbative expansion should be viewed
as a sort of sum rule that should be satisfied by a detailed calculation of the production spectrum.

In a recent publication~\cite{CMS:2025kzt}, a claim has been made on the observation of an excess
of the $t{\bar t}$ production cross section for small $M_{t{\bar t}}$ compatible with
the production of an $\eta_t$ pseudoscalar resonance.
In our view, the question of whether such an excess can be unambiguously attributed to a bound state
resonance
is a delicate one, since there are several sources of threshold enhancement and depletion, and they
should all be considered. There is nearly no doubt that the $\eta_t$ resonance must be present
in the production spectrum, and that with a fine resolution on the measurement of $M_{t{\bar t}}$
 it should become visible. However, when smearing the production cross
section with the resolution that is currently available at the LHC, several other contributions
come into place. First of all, $\as/v$ and $(\as/v)^2$ effects yield an important
enhancement to the cross section. Cubic corrections of order $\as^3 \delta(M_{t{\bar t}}-2m)$, unrelated
to bound state formation, can deplete the signal by a factor of 2. Other subtle QCD effects come
into place that are responsible for the reduction of the strong coupling constant relevant for
cross sections that are smeared out over the threshold region, as we find in our work.
It will be interesting to apply our analysis to the CMS results to clarify these points.

As already stated in the introduction, we are not the first authors to
show that one should not add bound state contributions to the
perturbative expansion for integrated cross sections. As can be seen
by looking at the previous literature~\cite{Beneke:2016jpx, Melnikov:2014lwa,
  Braun:1968njz, Novikov:1977dq, Voloshin:1979uv, Voloshin:1984zzn, Smith:1994ev}, however, it seems that this
message has not yet become common knowledge in the theory community,
so that in several different contexts (including the present one)
researchers have stumbled on this issue. We hope that the
simple derivation that we presented here will help in making this message stick
as common knowledge in the theoretical physics community.

\section*{Acknowledgements}
We wish to thank Fabio Maltoni and Pier Monni for useful conversation,
Davide Pagani for useful conversation and comments on the paper,
Martin Beneke for pointing out ref.~\cite{Beneke:2016jpx} to us and
for useful comments on the paper, and Kirill Melnikov for making us
aware of refs.~\cite{Melnikov:2014lwa,Braun:1968njz}.  P.N. would like
to thank Orfeo Nason for a useful discussion regarding the issue
illustrated in Section~\ref{sec:onehalf}.

The work of E.R. is partially supported by the Italian Ministero
dell'Universit\`a e Ricerca (MUR) through the research grant
20229KEFAM (PRIN2022, Finanziato dall'Unione europea - Next Generation
EU, Missione 4, Componente 1, CUP H53D23000980006). This work is also
partially supported by ICSC - Centro Nazionale di Ricerca in High
Performance Computing, Big Data and Quantum Computing, funded by
European Union - NextGenerationEU.

\appendix

\section{Detailed calculation for the $t{\bar t}$ system}
\label{app:realcase}
In analogy with the model of Section~\ref{sec:toymodel} we consider the integration of the $\rho({\cal E})$ in eq.~(\ref{eq:ttrho}) multiplied by a power of the energy. We now examine the integral of the second term of
eq.~(\ref{eq:ttrho})
\begin{equation}
  I_n=\int_0^{{\cal E}_{\rm cut}}\mathd{\cal E} \, {\cal E}^n
 \frac{1}{4\pi^2}(m)^{3/2} \sqrt{\cal E}
 F\left(\frac{b_l\sqrt{m}}{\sqrt{\cal E}}\right) \,,
\end{equation}
and rewrite it in the following form
\begin{eqnarray}
I_n&=&\frac{b_lm^2}{4\pi^2} \int_0^{{\cal E}_{\rm cut}}\mathd{\cal E} \, {\cal E}^n
\left[\frac{1}{1-\exp(-\frac{b_l\sqrt{m}}{\sqrt{\cal E}})}
-\sum_{i=0}^{2n+1} f_i\left(-\frac{b_l\sqrt{m}}{\sqrt{\cal E}}\right)^i \right]
\nonumber \\
&+& \frac{b_l m^2}{4\pi^2} \int_0^{{\cal E}_{\rm cut}}\mathd{\cal E} \, {\cal E}^n
\sum_{i=0}^{2n+1} f_i\left(-\frac{b_l\sqrt{m}}{\sqrt{\cal E}}\right)^i,
\label{eq:appsubadd}
\end{eqnarray}
where the coefficients $f_i$ are defined by the equation
\begin{equation}
\frac{1}{1-\exp(z)}=\sum_{i=0}^\infty f_i z^i\,.
\end{equation}
The subtracted term in the square bracket of
eq.~(\ref{eq:appsubadd}) is integrable for ${\cal E}\to 0$, since its
most singular term behaves as ${\cal E}^{-(2n+1)/2}$, and it is
multiplied by ${\cal E}^n$. Furthermore, if we expand it
for large ${\cal E}$, the leading
term of the expansion behaves as ${\cal E}^{-(2n+3)/2}$, since the term of order
${\cal E}^{-(2n+2)/2}$ is missing.  This follows from the identity
\begin{equation}\label{eq:oddness}
  \frac{1}{1-\exp(-z)}+  \frac{1}{1-\exp(z)}= 1,
\end{equation}
which implies that for even $i>0$ $f_i=0$.
Thus the integral is convergent
also for ${\cal E}_{\rm cut}\to \infty$,
and we can separate
\begin{equation}
I_n=I_n^{(1)}+I_n^{(2)}+I_n^{(3)}
\end{equation}
where
\begin{eqnarray}
I_n^{(1)}&=&\frac{b_lm^2}{4\pi^2} \int_0^\infty \mathd{\cal E} {\cal E}^n
\left[\frac{1}{1-\exp\left(-\frac{b_l\sqrt{m}}{\sqrt{\cal E}}\right)}
             -\sum_{i=0}^{2n+1} f_i\left(-\frac{b_l\sqrt{m}}{\sqrt{\cal E}}\right)^i \right]
  \label{eq:I1}\\
I_n^{(2)}&=&-\frac{b_lm^2}{4\pi^2} \int_{{\cal E}_{\rm cut}}^\infty \mathd{\cal E} {\cal E}^n
             \left[\frac{1}{1-\exp\left(-\frac{b_l\sqrt{m}}{\sqrt{\cal E}}\right)}
             -\sum_{i=0}^{2n+1} f_i\left(-\frac{b_l\sqrt{m}}{\sqrt{\cal E}}\right)^i \right]\\
  I_n^{(3)} &=&\frac{b_lm^2}{4\pi^2} \int_0^{{\cal E}_{\rm cut}} \mathd{\cal E} {\cal E}^n
\left[
\sum_{i=0}^{2n+1} f_i\left(-\frac{b_l\sqrt{m}}{\sqrt{\cal E}}\right)^i \right].
\end{eqnarray}
In $I^{(2)}$ the range of integration requires ${\cal E} > {\cal E}_{\rm cut}$,
and thus the integrand can be expanded into a convergent series
in powers of $b_l\sqrt{m}{\sqrt{\cal E}}$, so that we can write
\begin{equation}
  I^{(2)}=-\frac{b_lm^2}{4\pi^2} \int_{{\cal E}_{\rm cut}}^\infty \mathd{\cal E} {\cal E}^n
  \left[
    \sum_{i=2n+3}^\infty
    f_i\left(-\frac{b_l\sqrt{m}}{\sqrt{\cal E}}\right)^i \right]\,.
\end{equation}
It is now clear that we can write
\begin{equation}\label{eq:I2I3}
  I^{(2))}_n+I^{(3))}_n=\int_0^{{\cal E}_{\rm cut}}\mathd{\cal E} \, {\cal E}^n
  \frac{1}{4\pi^2}(m)^{3/2} \sqrt{\cal E}
 F_+\left(-\frac{b_l\sqrt{m}}{\sqrt{\cal E}}\right),
\end{equation}
where with $F_+$ we indicate the formal power expansion of $F$, where the
(singular) coefficients are meant to be computed by analytic regularization.

The $I^{(1)}$ integral is finite, and thus we can choose to regularise it in any
way we like. We choose analytic regularization, that is achieved by allowing
$n$ to be non-integer, and at the end to continue $n$ to integer values.
The sum in the square bracket of eq.~(\ref{eq:I1}) leads to vanishing scaleless integrals,
and we are left with
\begin{equation}
  I^{(1)}_n=\frac{b_lm^2}{4\pi^2}\int_0^\infty \mathd {\cal E}\frac{{\cal E}^n}
  {1-\exp\left(-\frac{b_l\sqrt{m}}{\sqrt{\cal E}}\right)}.
\end{equation}
The integration procedure depends upon the sign of $b$, but we can use eq.~(\ref{eq:oddness}) to derive the relation
\begin{equation}
  \frac{1}
  {1-\exp\left(-\frac{b_l\sqrt{m}}{\sqrt{\cal E}}\right)}
    =\frac{-|b_l|}{b_l} \frac{1}
  {1-\exp\left(\frac{|b_l|\sqrt{m}}{\sqrt{\cal E}}\right)}+\theta(b_l)\,,
\end{equation}
where the $\theta(b_l)$ term yields a vanishing contribution to the analytically
regulated integral,
so that
\begin{eqnarray}
  I^{(1)}_n
  &=&\frac{-b_l}{|b_l|}\frac{b_lm^2}{4\pi^2}\int_0^\infty \frac{{\cal E}^n}
               {1-\exp\left(\frac{|b_l|\sqrt{m}}{\sqrt{\cal E}}\right)}
      = \frac{m^2|b_l|}{4\pi^2}(mb_l^2)^{n+1}
      \int_0^\infty \mathd x \frac{x^n}{\exp(x^{-1/2})-1}
      \nonumber \\
  &=&\frac{m^2|b_l|}{4\pi^2}(mb_l^2)^{n+1}\int_0^\infty \mathd z \frac{2z^{-2n-3}}{e^z-1}
      =\frac{m^2|b_l|}{4\pi^2}(mb_l^2)^{n+1}
      2\Gamma(-2n-2)\zeta(-2n-2).\phantom{aaaaa}
\end{eqnarray}
Using the functional equation
\begin{equation}
  \zeta(s)=2^s\pi^{s-1} \sin\frac{\pi s}{2} \Gamma(1-s)\zeta(1-s)
\end{equation}
we obtain
\begin{equation}
  \Gamma(-2n-2)\zeta(-2n-2)=\frac{\zeta(2n+3)}{2^{2n+3}\pi^{2n+2} \sin(n\pi+3\pi/2)}
  \to (-)^{n+1} \frac{\zeta(2n+3)}{2^{2n+3}\pi^{2n+2}}
\end{equation}
where the arrow indicates the limit for integer $n$. Setting $b=\pi a$ we finally obtain
\begin{equation}
  I^{(1)}_n=\frac{|a| m^2}{4\pi}\left(-\frac{ma^2}{4}\right)^{n+1}\zeta(2n+3).
\end{equation}
This turns out to be identical to the form
\begin{equation}\label{eq:I1last}
  I^{(1)}_n=\int_0^\infty \mathd {\cal E}{\cal E}^n \left[\frac{-a_l}{|a_l|}
    \frac{1}{2\pi r_l^3}\sum_{j=1}^\infty \frac{1}{j^3}\delta({\cal E}-E_{l,j})\right].
\end{equation}
Using the identiy
\begin{equation}
  \theta(a) - \frac{a}{2|a|} = \frac{1}{2}, 
\end{equation}
we can combine eqs.~(\ref{eq:ttrho}), (\ref{eq:I2I3}) and (\ref{eq:I1last})
to obtain eq.~(\ref{eq:rhoFull}).
\bibliographystyle{JHEP}

\bibliography{ttbcorr.bib}
 
\end{document}